\documentclass{interact}

\usepackage{natbib}
\bibpunct[, ]{(}{)}{;}{a}{}{,}
\usepackage[noperiod]{jabbrv}

\usepackage{amssymb,amsbsy}
\usepackage{amsmath}
\usepackage{dsfont}
\usepackage{graphicx}
\usepackage[dvipsnames]{xcolor}
\usepackage{hyperref}
\usepackage[capitalise]{cleveref}
\usepackage[normalem]{ulem}  
\usepackage{geometry}
\usepackage{subcaption}
\usepackage{cite}
\usepackage{alphalph}

\newcommand{\var}[1]{\mathbb{V} \left( {#1} \right)}
\newcommand{\relvar}[1]{\mathbb{V}_r \left( {#1} \right)}
\newcommand{\urelvar}[1]{\mathbb{V}_{ur} \left( {#1} \right)}
\newcommand{\esp}[1]{\mathbb{E}\left[ {#1} \right]}

\newcommand{\varw}[1]{\mathbb{V} \left( {#1} \right)}
\newcommand{\relvarw}[1]{\mathbb{V}_r \left( {#1} \right)}
\newcommand{\espw}[1]{\mathbb{E}\left[ {#1} \right]}

\begin{document}          

\title{A Variance-Decomposition Formula for Direct and Adjoint Monte Carlo Particle Transport Problems}

\author{
\name{P. Rovel\textsuperscript{a,b}\thanks{CONTACT P. Rovel. Email: paul.rovel@cea.fr}, D. Mancusi\textsuperscript{a}, C. Larmier\textsuperscript{a} and A. Zoia\textsuperscript{a}}
\affil{\textsuperscript{a}Université Paris-Saclay, CEA, Service d'études des réacteurs et de mathématiques appliquées 91191 Gif-sur-Yvette, France; \textsuperscript{b}École nationale des ponts et chaussées, Institut Polytechnique de Paris, 77455 Champs-sur-Marne, France}
}

\maketitle

\begin{abstract}
    In fixed-source Monte Carlo particle-transport problems, obtaining an acceptable variance on the sought response is of paramount importance. Currently, the only rigorous tool to analyze the variance of such games is the framework of the moment equations, which is unfortunately unwieldy to use in practice. In this paper, we establish a formula that enables expressing the variance of a Monte Carlo simulation as a sum of `variance contributions' collected throughout the underlying particle-transport process. This formula applies to both direct and adjoint games, and provides a new tool to pinpoint the variance-inducing mechanisms in Monte Carlo simulations, understand common variance-reduction techniques, and even conceive new ones. We showcase the use of the variance-decomposition formula on several applications. In particular, we revisit zero-variance schemes and analyze existing variance-reduction techniques, underlining their strengths and weaknesses. A few relevant numerical examples substantiate our theoretical findings.
\end{abstract}


\section{Introduction}

When solving fixed-source transport problems using Monte Carlo methods, the variance $\sigma^2$ of the average sample of the sought observable, say the response function at a given detector, is a key quantity to monitor in the course of the simulation \citep{lux_monte_2018}. The variance determines the potential influence of stochastic fluctuations on the obtained result and, assuming a Gaussian distribution, can be converted into a confidence interval. Since $\sigma^2$ decreases proportionally to the number of (independent) simulated histories, the efficiency of a given Monte Carlo scheme is conventionally expressed by the Figure of Merit (FoM) \citep{lux_monte_2018}, defined as
\begin{equation}
    \text{FoM}= \frac{1}{\sigma^2 t},
\end{equation}
where $t$ is the total simulation time. Improving the FoM is an essential goal for Monte Carlo practitioners: a larger FoM implies better statistical convergence for the same computer resources, or equivalently less computation time for the same statistical accuracy. 

In some Monte Carlo simulations, for example deep-penetration problems, achieving reasonable FoM without so-called `variance-reduction techniques' might turn out to be simply infeasible. Variance-reduction techniques are very diverse, and act upon very different aspects of the Monte Carlo sampling process. A non-exhaustive list of such method includes e.g.~\emph{implicit absorption}, which forbids the sampling of absorption reactions; \emph{path stretching}, which favors the sampling of flights oriented towards the detector region; \emph{Weight Windows} \citep{booth_genesis_2006}, which make use of Russian roulette and splitting to increase the number of particles approaching the detector; \emph{branchless collisions} \citep{lux_monte_2018}, which forbids branching events by sampling reaction paths proportionally to their multiplicities and keeping the number of emitted particles at each collision equal to one; \emph{importance sampling}, which consists in sampling the flights and collisions proportionally to a suitably-chosen importance function; and \emph{Consistent Adjoint-Driven Importance Sampling (CADIS)}\citep{wagner_automated_1998}, which makes use of a deterministic adjoint calculation to generate an importance map used in both source sampling and Weight Windows.

These techniques have mostly been derived empirically, and can usually be justified by an intuitive reasoning: for instance, it seems reasonable that pushing particles towards the detector will help in having more contributions within the tallying region, thus yielding a better statistical convergence.

There exist a few tools to rigorously explain how these techniques affect the final variance, and to describe why they may possible fail in some cases. Perhaps the most powerful theoretical framework to interpret the behavior of variance-reduction methods is the moment equations formalism \citep{lux_monte_2018}, which is in particular capable of determining the variance of the tally associated to a particle of unit weight starting at a given point $P$ in phase space. This approach provides an integral equation for the variance as a function of the different processes simulated during the Monte Carlo game. This tool has been used, for example, to derive the rules of zero-variance games, by formally equating the variance to zero and finding the sampling rules that satisfy this condition. However, the integral equation for the second moment is unwieldy to use in practice: not only solving it is generally more complicated than solving the actual problem, i.e.~the equation for the first moment, but interpreting the effects of a given variance-reduction technique on its terms is usually difficult. This approach has thus rather limited explanatory capacity.

In this paper, we focus on the variance components of the FoM, and provide a formula expressing the final variance of a fixed-source Monte Carlo calculation in terms of a sum of variance contributions added by the different sampling events executed during the Monte Carlo game. Similarly to the integral equation for the second moment, the terms composing the sum in the formula proposed in this work are not promptly obtained, but they are simpler to wield and allow for an easy interpretation of most variance-reduction techniques. Furthermore, the variance-decomposition formula is expressed in a general framework, and can be thus generalized to other Monte Carlo problems (not necessarily related to transport). In the scope of this paper, the main requirement for the variance-decomposition formula to apply is that the game is `positive', meaning the statistical weights remain positive and the particle can only contribute positive quantities (this hypothesis could be lifted to obtain a generalized formula but this is outside the scope of this work). An example of Monte-Carlo game not related to transport that the variance-decomposition formula can handle is the evaluation of the integral of a function via Monte-Carlo \citep[see][Chap 4, Part I]{lux_monte_2018}, provided that the function is always positive.

This paper is organized as follows. In Sec.~\ref{sec:formula} we introduce the variance-decomposition formula and detail in an intuitive way the structure of all the terms appearing therein. A simple expression approximating the calculation time is also derived using the same terms, yielding a global decomposition formula for the FoM. For the sake of readability, the mathematical derivation of the formula is provided in \cref{sec:proof}. The \cref{sec:variance_reduction_techniques,sec:zero_variance,sec:examples} focus on a few relevant applications of the variance-decomposition formula. In \cref{sec:variance_reduction_techniques}, we show how common variance-reduction techniques such as Importance Sampling, implicit capture, branchless, path stretching, and CADIS can be interpreted and decomposed using the proposed formula. The question of how to distribute particles in phase space (via Weight Windows, for example) to optimize both variance reduction and calculation time is answered within a FoM-optimal approach. In \cref{sec:zero_variance}, the formula is then used to derive the general strategy of zero-variance games for direct and adjoint transport problems, which allows recovering the known zero-variance expressions for the sampling process. Finally, in \cref{sec:examples}, the variance-decomposition formula is evaluated for several test-cases. We first consider a benchmark problem admitting an analytical solution, which allows comparing the exact expressions of the moment equations and the variance-decomposition formula. Then, we numerically solve some direct and adjoint particle-transport problems and illustrate for each the behavior of the formula. Technical details are collected in a series of Appendixes.

\section{The variance-decomposition formula}
\label{sec:formula}

\subsection{Context and definitions}
\label{sec:context_and_def}

Before introducing the variance-decomposition formula, we will discuss the general setting in which it can be applied. We will show that several well-known Monte Carlo problems, encompassing direct and adjoint particle transport and integral evaluation, fit into that general setting.

We consider our Monte Carlo game to be a (possibly branching) Markov process occurring on a phase space denoted $\Pi$; we will denote $P$ the points of that phase space. $\Pi$ can be continuous, discrete or a mixture of both. The game contains an ensemble of particles\footnote{We use the term `particle', which stems from transport problems, for the sake of readability. More broadly, we could refer to `walkers'.}. Each particle is endowed with a statistical weight, denoted $w$. A fixed number of particles, denoted $N$, begin the game at a given starting point in $\Pi$ that we will denote $P_\text{source}$, with an initial weight set to unity. 

The game progresses by successive stochastic sampling events that act upon a single particle (provided that it has not been terminated). The collection of these events defines the `history' of the particle. A particle currently at point $P$ will undergo a stochastic sampling event that can either terminate the current history by sending the particle to a specific point denoted $P_\text{death}$, or create a number $K\ge 1$ of offsprings at points $P'_k$.  The $K$ offsprings will be indexed by $k\in[1,K]$, and their weights $w'_k$ are determined by applying a multiplier $m_k\ge0$ on the initial particle weight $w$: $w'_k=m_kw$. Since $m_k\ge0$ and the initial weights are all positive, all the weights remain positive. The Markov nature of the game imposes that all the stochastic events that act upon particles starting from a given point $P$ are independent and identically distributed (iid). All the sampling events starting at point $P$ follow the law of the random variable 
\begin{equation}
    X_P:=(K,P'_1,m_1,...,P'_k,m_k).
    \label{eq:def_xp}
\end{equation}
For the sake of conciseness, we will refer to $X_P$ as `the sampling from point $P$'. The multiplicative nature of the weight modification (via the multipliers $m_k$) and the independence of the different particles impose a form of linearity with respect to the average total weight entering at point $P$; this will be detailed below. A particle being in $P_\text{death}$ is removed from the game; the game stops when no particles remain. 

Special sampling events that depend on the entering weight $w$ of the particle (e.g. a Russian roulette procedure, whose survival probability depends on the entering weight) cannot be treated directly within the framework. However, they can be taken into account provided that they satisfy the linearity with respect to the average total weight (which is the case of weight control procedures, in order to preserve unbiasedness). To circumvent the problem, we might adapt the phase space $\Pi$, so that each weight has a different entry point $P_w$ for those weight-dependent events, effectively creating different samplings $X_{P_w}$ for different weights. Another solution would be to separate all points $P$ in $\Pi$ based on the weight $w$. This latter strategy would keep the variance decomposition formula valid, but it would make it less easy to manipulate. In the remainder of this work, we choose the first strategy: include the weights in the phase space points only for the specific weight-dependent sampling events. The use and interpretation of the variance decomposition formula for weight correction procedures (that are weight dependent) is treated in depth in \cref{sec:pop_control}.   

Throughout the game, a score is collected based on the particles' histories. When a particle lands on point $P$ with a weight $w$ it contributes $h(P)w$ to the score, where $h$ is called the estimator function, which is supposed to be non-negative everywhere ($h(P)\ge0$). We define the random variable $R$ as the total score collected during the game by all particles, divided by the number $N$ of source particles. The quantity $R$ will be referred to as the `result' of the Monte Carlo game; the associated quantity $\esp{R}$ represents thus the `sought response' of the game. As all weights are non-negative and the response function is non-negative, contributions are non-negative and $R$ is non-negative. We will say that the game is `positive' if $m_k\ge0$ and $h\ge0$. This hypothesis could be lifted, yielding a generalized version of the variance decomposition formula, but this is outside the scope of this work.

We introduce the particle density denoted $n$, which is a distribution\footnote{We assume that $n(P)dP$ is a positive finite measure.} over $\Pi$. The quantity $n(P)dP$ corresponds to the average number of particles (irrespective of their weights) in $dP$ around $P$. Note that at discrete points such as $P_\text{source}$, where particles have a non-zero probability of being exactly at that point, the density $n$ uses the discrete measure and simply represents the average number of particles at the point. Additionally, we introduce the matter density $\rho$ such that $\rho(P)dP$  corresponds to the sum of the statistical weights of particles in $dP$ around $P$, normalized by the number $N$ of source particles. Note that $\rho$ takes the particle weights into account, contrary to $n$.

We introduce then the moment-related quantities, which are functions of $\Pi$. The $i$\textsuperscript{th} moment with respect to the estimator function $h$ at point $P$, denoted $M_i(P)$, is the expectation of the $i$\textsuperscript{th}-power of the contribution to the score associated to a single particle starting with unit weight at point $P$. Denoting $R_P$ the random variable describing the result (estimated using $h$) of a game where only one source particle is placed with unit weight at point $P$, we have:
\begin{equation}
    M_i(P)= \esp{(R_P)^i}.
    \label{eq:def_ith_moment}
\end{equation}
As all contributions are positive, the moments are also positive. The first moment $M_1$ is called the importance function.

Note that the sampling structure introduced above, which acts independently on all particles and applies weight multipliers $m_k$, naturally behaves linearly with respect both to the matter density $\rho$ and to the first moment $M_1$. This means that the expected contribution of $l$ particles of weight $w$ at point $P$ is $l\times w \times M_1(P)$. Similarly, the density $\rho$ induced by $l$ particles of weight $w$ is $lw$ times the density $\rho$ induced by a single particle of unit weight.

The general framework presented above is well-suited to direct particle-transport problems. The sampling events $X_P$ will represent all the sampling procedures happening in the course of Monte Carlo transport simulation, e.g. source, flights and collisions. The phase space $\Pi$ acts as a `generalized phase space', where points $P$ include the standard phase-space point $\mathcal{P}=(\mathbf{r},E,\mathbf{\Omega})$ required by the Boltzmann equation, plus the current state of the particle (e.g. being emitted or entering a collision), so that the next sampling to perform is entirely defined by $P$.

There is some leeway in the choice of what is represented by $X_P$: we could e.g. choose to represent a whole cycle flight+collision as a single sampling $X_P$, in which case $\Pi$ would only contain points corresponding to particles being emitted; conversely, we could choose to decompose the sampling of collisions into sub-samplings (e.g. sampling of the nuclide $i$, sampling of the reaction $j$ within the nuclide $i$, sampling the outgoing distribution for reaction $j$ in nuclide $i$), in which case the point $P$ would include more detailed particle states. The point $P_\text{source}$ corresponds to the state of a particle before it has been sampled by the source of the problem; correspondingly, $X_{P_\text{source}}$ corresponds to the sampling of the source.

The use of an estimator function $h(P)$ would naturally correspond to the use of collision estimators for particle-transport problems, where $h$ is non zero only if $P$ corresponds to a point of entering collision. Throughout this paper, we will consider that transport problems use collision estimators; however, this can straightforwardly be generalized to estimators that depend on transitions between points (such as track-length estimators). This is done in \cref{sec:non_col_estimators}. The matter density $\rho(P)$ then corresponds to the collision density $\psi(\mathcal{P})$ or the emission density $\chi(P)$, depending on whether $P$ represents a collision or an emission point. The first moment $M_1(P)$ corresponds to the emission importance $\chi^\dagger(P)$ or the collision importance $\psi^\dagger(P)$, that can be obtained by solving the adjoint Boltzmann equation, depending if $P$ represents an emission point or collision point.

Unsurprisingly, this framework also applies to adjoint particle-transport problems by simply switching the roles of the importance and the collision and emission density. The matter density $\rho$ would correspond to either $\psi^\dagger$ or $\chi^\dagger$, and the first moment $M_1(P)$ would correspond to $\psi(P)$ or $\chi(P)$.

Finally, this framework also applies to the evaluation of integrals of functions $f$ over a domain $\mathcal{D}$ using Monte Carlo sampling. The usual strategy consists in sampling $N$ random points $x_n$ in $\mathcal{D}$ distributed according to a distribution $g$. The estimation of $\int_\mathcal{D}f(x)dx$ is obtained by computing the average over the $N$ samples of $f(x_n)/g(x_n)$. In this case, our phase space is $\Pi = \{ P_\text{source},P_\text{death}\} \cup \mathcal{D}$. The first sampling $X_{P_\text{source}}$ corresponds to sampling $P_1'$ in $\mathcal{D}$ with $K=1$ and $m_1=1$. At point $P = x\in \mathcal{D}$, particles score $h(P')=f(x)/g(x)$, and go to $P_\text{death}$.

\subsection{Introducing the variance-decomposition formula}

We can now introduce the variance-decomposition formula, give a concise sketch of the proof, and provide physical intuition concerning all the terms occurring therein. The formal proof of the formula and the rigorous definition of the involved terms is deferred to \cref{sec:proof}. 

Contrary to most works on the variance of a Monte Carlo game, our derivation is not based on the second moment equation. The proof is based instead on the law of total variance, a foundational theorem in probability theory that is used to `decompose' the variance of a random variable $A$ into a sum of terms representing the `contributions' to the variance of $A$ due to a series of other random variables;  \citep[see e.g.][chapter 4, section 4]{casella_statistical_2024}. For our goals, the law of total variance is applied to $R$, to `decompose' its variance based on the different samplings occurring in chronological order during the Monte Carlo simulation. Since interpreting the nature of the $i^\text{th}$ sampling in chronological order can be difficult, the decomposition is re-arranged based on the point $P$ where the current particle is when the $i^\text{th}$ sampling occurs. This yields the variance of $R$ expressed as an integral over all different points in $\Pi$. The integrand is then manipulated to obtain terms that are physically interpretable.

The variance-decomposition formula expresses the final relative variance\footnote{The relative variance $\mathbb{V}_r$ is the variance divided by the squared expected value: $\relvar{X}=\var{X}/\esp{X}^2$.} $\relvar{R}$ of the obtained result $R$ as a sum of variance contributions that stem from the Monte Carlo sampling procedures in the phase space $\Pi$:
\begin{equation}
    \relvar{R} = \frac{1}{N}\int_\Pi c_r(P)\frac{1+\relvarw{w(P)}}{I_s(P)}\relvar{X_P} dP,
    \label{eq:decomp_var}
\end{equation}
where $c_r$ is the `contribution density', $\relvarw{w(P)}$ is the relative variance of the weights at point $P$, $I_s$ is the `sampling intensity', and $\relvar{X_P}$ is the `intrinsic variance of sampling' at point $P$. The terms occurring in \cref{eq:decomp_var} will be precisely defined in the following. 

Before introducing the `intrinsic variance' $\relvar{X_P}$ of the process $X_P$, we need to introduce the random variable describing the importance of the outcomes of sampling $X_P$, or simply `the importance of $X_P$':
\begin{equation}
    M_1(X_P) := h(P)+\sum_{k=1}^K m_k M_1(P'_k).
    \label{eq:def_importance_xp}
\end{equation}
The importance of $X_P$ takes into account the immediate payoff for a particle being in $P$ and adds the expected payoff due to the outcome of $X_P$. In the RHS of \cref{eq:def_importance_xp}, $h(P)$ is a deterministic payoff, since $P$ is fixed for a given $X_P$; the importance $M_1$ is a deterministic function, but its argument $P'_k$, $K$ and the $m_k$ are all random variables that depend directly on $X_P$ via \cref{eq:def_xp}. The `intrinsic variance' $\relvar{X_P}$ of the process $X_P$ is then a shorthand notation for:
\begin{equation}
    \relvar{X_P}:= \relvar{M_1(X_P)}= \var{\frac{\sum_{k=1}^K m_k M_1(P'_k)}{M_1(P)}},
    \label{eq:intrinsic_variance}
\end{equation}
where we used the fact that $\esp{M_1(X_P)}=M_1(P)$, and we removed the constant $h(P)$ term from the numerator.

The $\relvar{X_P}$ term in \cref{eq:decomp_var} is weighted by terms representing how important the sampling $X_P$ is with respect to the result $R$. The relative contribution density $c_r(P)$ at point $P$ is a distribution\footnote{In the sense that $c_r(P)dP$ is a positive finite measure.} representing the fraction of the total response $\esp{R}$ that will pass in $dP$ around point $P$ on average:
\begin{equation}
    c_r(P):= \frac{\rho(P)M_1(P)}{\esp{R}}.
    \label{eq:def_rel_contrib}
\end{equation}
Observe that $c_r(P)$ corresponds to the quantity defined as the `response density' in the contributon theory proposed in \citet{williams_generalized_1991}. For example, if the point $P$ corresponds to the state of a particle being emitted, the relative contribution would be $c_r(P) = \chi(\mathcal{P})\chi^\dagger(\mathcal{P})/\esp{R}$, with $\chi$ the emission density and $\chi^\dagger$ the adjoint emission density (i.e.~the importance of an emitted particle). One of the main properties of the contribution density is that it describes a flow that emerges in the source and exits in the detector. The total response $\esp{R}$ can then be seen as a conserved quantity that can be measured by integrating the contribution on any surface that stands between the source and the detector in $\Pi$. See \citet{williams_generalized_1991} for more details regarding the contribution density.

Two other weighting terms are applied to the `intrinsic variance' $\relvar{X_P}$ in \cref{eq:decomp_var}, acting as variance-amplification factors. The first is the relative variance of the weights $\relvarw{w(P)}$, which describes how dispersed the statistical weights are at point $P$, i.e.~right before undergoing the sampling $X_P$. Note that the notion of average weight at point $P$ (denoted $\espw{w(P)}$) and the quantities that follow (average square of the weights $\espw{w^2(P)}$, variance of the weights $\varw{w(P)}$ and relative variance of the weights $\relvarw{w(P)}$) need precise definitions, as the weight at point $P$ is not strictly speaking a random variable: in some realizations of the game, e.g. particles might never visit point $P$, while in others several particles might visit $P$. Therefore, we define the average weight at point $P$ as the average sum of the weights of particles at $P$ (corresponding to $N\rho(P)$), divided by the average number of particles $n$ at $P$:
\begin{equation}
    \espw{w(P)} := \frac{\rho(P)}{n(P)/N}.
    \label{eq:def_avg_weight_from_rho}
\end{equation}
The quantity $\espw{w^2(P)}$ is defined similarly as the sum of the squared weights at point $P$, divided by $n(P)$. The variance of the weights $\varw{w}$ is then defined as $\espw{w^2}-\espw{w}^2$ and the relative variance of the weights$\relvarw{w}$ as $\varw{w}/\espw{w}^2$. We do use the notations $\mathbb{E}$ and $\mathbb{V}$ as these quantities do correspond to actual expected values and variances in a different probability space (this is discussed in depth in \cref{sec:proof}).

The second weighting term $I_s$ represents the  `sampling intensity', and describes how many particles (on average) go through the sampling $X_P$, compared to how important the sampling $X_P$ is to the result $R$. It is defined as:
\begin{equation}
    I_s(P) = \frac{n(P)}{Nc_r(P)},
    \label{eq:sampling_intensity}
\end{equation}
i.e.~the fraction of particles at point $P$ divided by the fraction of the total contribution at point $P$. For instance, a region where half the response $\esp{R}$ flows on average will have a sampling intensity of one if half the particles pass in this region on average. A sampling intensity larger than one means that we `over-sample' the region, and smaller than one means that we `under-sample' the region.

To conclude, \cref{eq:decomp_var} is thus a weighted integral collecting the contribution to the relative variance $\relvar{R}$ of the sampling $X_P$ over the different points $P$ of phase space. If the Monte Carlo game is not a particle-transport problem, the generalized phase space can take different shapes. In \cref{sec:examples}, e.g.~we will consider a Monte Carlo game based on a discrete phase space; in such an example, the integral over the phase space $\Pi$ occurring in \cref{eq:decomp_var} will use the discrete measure and turn into a sum over the different states.

\subsection{Interpretation of the formula}
\label{sec:interp_var_decomp}

The variance-decomposition formula in the form given in \cref{eq:decomp_var}, albeit formal, can be usefully given a physical interpretation. 

First, the final relative variance has a $1/N$ term, $N$ being the number of source particles. As the $N$ source particles are independent (all the samplings $X_P$ are `Markov'), the Monte Carlo game can be extended to any number $N$ of source particles. The terms within the integral in \cref{eq:decomp_var} being identical for any $N$, we recover the expected $1/N$ behavior of the variance.

The final variance is a weighted integral of intrinsic variances $\relvar{X_P}$ of the different sampling processes $X_P$ occurring in the Monte Carlo game. The variance of a sampling process $X_P$ is quantified by the variance of the importance $M_1(X_P)$ of its outputs, as shown in \cref{eq:intrinsic_variance}.

Of course, the intrinsic variance terms at different points $P$ will not have the same influence on the final variance, which is why they need to be weighted first. In \cref{eq:decomp_var}, the intrinsic variances are weighted by $c_r(P)dP$, i.e. the fraction of the total contribution that passes through the sampling starting in $dP$ around point $P$. Hence, a sampling process with a large intrinsic variance, but located in a region that contributes little to the sought response (e.g. in a region far away from both the source and the detector region), will have a small relative contribution going through it and thus will not significantly impact the final variance.

The sampling intensity $I_s$ describes how intensely the zone $P$ is sampled compared to its relative contribution. If a point $P$ receives on average a fraction of particles that is more important than its relative contribution, $I_s(P)>1$, then the zone is `over-sampled', and the impact of the intrinsic variance on the final variance will be reduced. On the other hand, if the zone is `under-sampled', then the impact of the intrinsic variance will be exacerbated.

Finally, the term $1+\relvarw{w(P)}$ conveys the fact that, if there is dispersion of the statistical weights of the particles that enter the sampling $X_P$ (relative to the average weight observed entering sampling $X_P$), then the initial dispersion will also amplify the intrinsic variance of the process.

Let us now detail upon which term one can act to reduce the variance of his Monte Carlo calculation. If one wishes to obtain the sought result $\esp{R}$, the modification applied needs to keep the game unbiased. We will restrict ourselves to modifications that keep the source point $P_\text{source}$, the estimator $h$\footnote{Regarding transport problems, we restricted ourselves to collision estimators. In \cref{sec:non_col_estimators} where we treat all estimators, we will treat the question of changing the estimator to another partially unbiased estimator.}, the matter density $\rho$ and the importance $M_1$ unchanged. Lifting these requirements would authorize pathological games, such as the game where one scores $\esp{R}$ at the source point and then no simulation is performed. This game is technically unbiased, but it is not practically feasible, as it requires knowing the solution to the problem beforehand. Preserving $\rho$ and $M_1$ implies that the relative contribution $c_r$ cannot be modified (see \cref{eq:def_rel_contrib}). In the framework of transport, this is expected, since $c_r$ can be computed using $\chi$ and $\chi^\dagger$ (respectively $\psi$ and $\psi^\dagger$), which only depend on the underlying Boltzmann transport equation and not on the Monte Carlo setting.

It is possible to alter the intrinsic variance of sampling $X_P$, by changing the probabilities of sampling the different outcomes of $X_P$. Modifying the probability distribution of $X_P$ implies modifying the weights multipliers $m_k$ accordingly, to ensure that $\rho$ and $M_1$ stay constant; this does not prevent us from acting on the $\relvar{X_P}$ term. Of course, modifying the sampling probabilities will also impact the distribution of particles in the phase space, and thus impact the sampling intensity $I_s$. Similarly, modifying the weight multipliers $m_k$ will impact the distribution of weights, and thus the relative variance of the weights $\relvarw{w(P)}$.

We can still influence the variance of the result $R$ even without modifying the samplings $X_P$: this is achieved by applying population control, splitting and rouletting particles. These weight operations technically add new sampling events $X_P$, but preserve the structure of all the other samplings. Weight operations can modify the density $n$ of particles in $\Pi$, affecting the term $I_s$, and modify the weights $w$, affecting the term $\relvarw{w(P)}$.

\subsection{Simulation time and FoM}

As stated in the introduction, the efficiency of a Monte Carlo game, as defined by the FoM, depends on both the variance $\relvar{R}$ of the result $R$ and the average simulation time $t$. In the previous sections, we have derived a formula for the relative variance $\relvar{R}$ of the game. In this section, we set out to derive a formula for the average simulation time $t$, based on a similar approach.

The exact calculation time of a Monte Carlo game depends on many implementation details. Here we will use the average total number of sampling events as an ersatz for the total amount of calculations, and thus for the actual average simulation time $t$. Using the notations introduced earlier, we can decompose all sampling events on the different points $P$ of phase space. From \cref{eq:sampling_intensity}, this yields the time-decomposition formula
\begin{equation}
    t \propto \int_\Pi n(P)dP =  N \int_\Pi c_r(P)I_s(P)dP.
    \label{eq:time_decomp}
\end{equation}

Combining then \cref{eq:time_decomp} and \cref{eq:decomp_var}, we can readily derive a formula for the FoM:
\begin{equation}
    (\text{FoM})^{-1} \propto \relvar{R}t \propto \left[ \int_\Pi c_r(P)I_s(P)dP\right]\left[ \int_\Pi c_r(P)\frac{1+\relvarw{w(P)}}{I_s(P)}\relvar{X_P}dP \right].
    \label{eq:FoM}
\end{equation}
From \cref{eq:FoM}, it is apparent that the number of source particles $N$ cancels out in the expression of the FoM, as expected. The replacement of the expected number of particles $n$ by the sampling intensity $I_s$ in \cref{eq:time_decomp} usefully underlines the ambivalent role played by the sampling intensity in variance reduction. On the one hand, $I_s$ occurs in the denominator of the variance-decomposition formula in \cref{eq:decomp_var}; on the other hand, $I_s$ also occurs as a multiplication factor in the time-decomposition formula in \cref{eq:time_decomp}. Furthermore, both terms appear as integrals over all sampling events in phase space, and the pre-factor $c_r(P)$ which cannot be modified by altering the simulation techniques, as shown in \cref{sec:interp_var_decomp}.

\section{Analysis of common variance-reduction techniques}
\label{sec:variance_reduction_techniques}

In this section, we will examine the behavior of a few well-known variance-reduction methods for transport problems in the light of the variance-decomposition formula. In the spirit of Monte Carlo practitioners, the goal of `variance-reduction' techniques, contrary to what their name suggests, is to improve the FoM of a game, rather than just reducing its variance. Although our work mainly focuses on the variance component of the FoM, in most cases we will be compelled to also consider the time component of the FoM. In the rest of this work, `variance reduction' will be then interpreted as `FoM improvement', unless explicitly otherwise stated.

\subsection{Implicit capture}

Implicit capture is one of the simplest variance reduction techniques, and consists in avoiding the reaction channels corresponding to capture. The remaining reaction channels will be sampled from
\begin{equation}
    \hat{\mathbb{P}}(\{i,j\} \text{ channel}) = \frac{N_i\sigma_{i,j}}{\Sigma_t-\Sigma_{c}},
\end{equation}
where $\Sigma_{c}$ is the total macroscopic capture cross section for all nuclides, and the applied weight correction will then be
\begin{equation}
    \hat{m} = \frac{\Sigma_t-\Sigma_{c}}{\Sigma_t}.
\end{equation}
As we use the collision estimator, this procedure is unbiased. Note that this would not have been true for estimators that count specifically the capture reactions, such as the last event estimator. Overall, we expect this technique to lead to variance reduction, since the captured particles do not contribute to the score (unless they already did) and, in some sense, `waste' the calculation time needed to simulate their lives. By forcing these particles to survive, we increase the chances that they will contribute to the result $R$, and thus improve variance.

We can in particular prove that implicit capture can only reduce the variance. The first effect of this method is that the number of particles after the capture process will be increased. This can only increase the sampling intensity term $I_s$ for subsequent sampling events. The term $I_s$ occurs at the denominator of the variance-decomposition formula: its increase will thus decrease the final variance. However, increasing the sampling intensity $I_s$ also means that we increase the computing time, as shown by \cref{eq:time_decomp}. Observe that the effects on computing time and on variance play against each other. We will show in the following that this is often the case when interpreting how variance-reduction techniques work in the light of the variance-decomposition formula. Furthermore, if one uses population-control techniques to ensure that the number of particles in different zones of phase space attains a particular value, then the sampling intensity $I_s$ is `fixed'. In this case, this first effect will disappear.

The second effect has a smaller impact on variance and does not affect the computing time. At each collision, we remove a possible outcome of the collision sampling, which reduces the intrinsic variance of this choice. For illustration, let us decompose the choice of the reaction in two parts: first we choose whether the reaction is capture or something else; then, if it is not capture, we choose the other reaction channels. Let us denote $\psi^{\dagger*}$ the importance of a collided particle conditioned on the fact that the selected reaction channel is not capture. The importance of sampling capture is zero (as an absorbed particle does not contribute to the tally), and the importance at collision before this choice is simply $\psi^\dagger$. The random variable $M_1(X_P)$ is then a `binary' random variable that can take two values: either $\psi^{\dagger*}$, with probability $p = (\Sigma_t-\Sigma_c)/\Sigma_t$, or $0$, with probability $q = 1- p = \Sigma_c/\Sigma_t$. The intrinsic variance of the choice of capture vs.~non-capture is then:
\begin{equation}
    \relvar{X_P} = \relvar{M_1(X_P)} = \frac{(\psi^{\dagger*}-0)^2pq}{(\psi^\dagger)^2}.
\end{equation}
If the probability $q$ of sampling capture and the probability $p$ of sampling something else are non-zero, and the importance $\psi^{\dagger *}$ of not sampling capture is also non-zero, then this sampling has an intrinsic variance that will contribute to the final variance. Thus, if we remove this component of variance by never sampling the capture, the contribution to the final variance will be smaller. Note that this effect does not depend on the sampling intensity. Therefore, it will still remain even if one uses population control to enforce the sampling intensity $I_s$. The detailed treatment of the use of population-control techniques will be discussed in \cref{sec:sampling_intensity_WW_Cadis}.

\subsection{Branchless collisions}

Branchless collisions are a sampling method that was devised to suppress the possibly branching nature of analog simulation. The main idea is to sample every reaction proportionally to its average multiplicity: instead of sampling according to the cross sections $\sigma_{i,j}$, we sample nuclide $i$, reaction $j$ with probability
\begin{equation}
     \frac{N_i\nu_{i,j}\sigma_{i,j}}{\sum_{i',j'} N_{i'} \nu_{i',j'}\sigma_{i',j'}} = \frac{\nu_{i,j}}{\nu_\text{avg}} \times \frac{N_i \sigma_{i,j}}{\Sigma_t},
     \label{eq:branchless_sampling}
\end{equation}
where we introduced the average multiplicity as 
\begin{equation}
    \nu_\text{avg} = \frac{\sum_{i',j'} N_{i'} \nu_{i',j'}\sigma_{i',j'}}{\Sigma_t}.
\end{equation}
A single particle is emitted, which implies a weight correction
\begin{equation}
    \hat{m}= \frac{\nu_\text{avg}}{\nu_{i,j}}
\label{eq:brancless_weight_correction}
\end{equation}
for the selection of the reaction. The multiplicity of the reaction $\nu_{i,j}$ is then applied by further multiplying by $\nu_{i,j}$, yielding a total weight multiplier of $\nu_\text{avg}$ for all reactions. With this technique, capture is never sampled, since its multiplicity is zero. In the case where there is only scattering (with a multiplicity of one) and capture, the branchless simulation is similar to implicit capture. Therefore, branchless simulation can be seen as a generalization of implicit capture to all reactions. 

By construction, branchless sampling forbids branching, and is thought to improve the FoM of the simulation \citep[see e.g.][Chapter V, Sec.~III.C]{lux_monte_2018}. A common argument for branchless simulation is that it reduces computing time by not creating branching histories. The number of outgoing particles in an analog simulation is $\nu_\text{avg}$, and $1$ in a branchless simulation. Thus, the decrease in the number of secondary particles is true if $\nu_\text{avg}>1$. However, having fewer particles means that the sampling intensity $I_s$ of all subsequent sampling events will be smaller, and the impact of their intrinsic variance on the global variance-decomposition sum will then be larger. As was the case with implicit capture, variance and computation time act in opposite directions. The effect of computing time reduction is not at all obvious on the FoM (see \cref{eq:FoM}): in some cases it might be beneficial, and in others it might be detrimental. This question will be treated more in depth in \cref{sec:sampling_intensity_WW_Cadis}. Furthermore, similarly to implicit capture, when using population control to enforce a certain population density in phase space the branching character of the collision has no impact on the number of particles after the collision.

However, like for implicit capture, there is a clear argument for branchless simulation, which holds even if population control is used. This is surprisingly not related to the `branchless' nature of the sampling. The fact that we sample each reaction taking into account its multiplicity is a rudimentary form of importance sampling, where the probability of each outcome is adjusted proportionally to its relative importance. To a first approximation, the importance of sampling a reaction is determined by the number of offsprings that each reaction will produce. Of course, since particles emerging from different reactions will have different distribution laws, their importance can vary for other reasons: for example fission produces high-neutrons that might have more importance than low-energy neutrons produced by inelastic scattering. However, supposing that particles emerging from a reaction are roughly equivalent, then importance sampling amounts to sampling proportionally to the multiplicity. Denoting $\chi^\dagger$ the importance (supposedly identical) of an offspring particle, the importance of a particle in channel $\{i,j\}$ is $M_1(\{i,j\} \text{ channel})=\nu_{i,j} \chi^\dagger$, and sampling according to \cref{eq:branchless_sampling} leads to weight multiplication as in \cref{eq:brancless_weight_correction}. Finally, the random variable $M_1(X_P)$ corresponding to the process of choosing a reaction is:
\begin{equation}
    \hat{m}M_1(\{i,j\} \text{ channel}) = \frac{\nu_\text{avg}}{\nu_{i,j}}\times \nu_{i,j}\chi^\dagger = \nu_\text{avg}\chi^\dagger= \text{constant},
\end{equation}
and the intrinsic variance of the process is zero. Reducing the intrinsic variance of every reaction choice will have a positive impact on the variance, independently of the fact that the resulting histories could be branching or not. One could in fact intentionally split the emitted particle into $\nu_\text{avg}$ outgoing particles to preserve particles of unit weight, and this would not impact the reduction of variance due to the importance sampling of the reaction choice.

\subsection{Sampling intensity, weight windows, and CADIS}
\label{sec:sampling_intensity_WW_Cadis}

We will now focus on the techniques known as Weight Windows \citep{booth_genesis_2006}. More generally, we will discuss the effect of the sampling intensity $I_s$ in the variance-decomposition formula given in \cref{eq:decomp_var}. This is arguably the most intricate term appearing in the formula, as its effect also directly impacts the calculation time (see \cref{eq:time_decomp}). Our analysis will then focus on both the variance and the calculation time.

\subsubsection{The extreme case: deep-penetration problems}

Let us start by considering a rather extreme case. The integrand in variance-decomposition formula in \cref{eq:decomp_var} for point $P$ reads
\begin{equation}
    c_r(P)\frac{1+\relvarw{w(P)}}{I_s(P)}\relvar{X_P}.
    \label{eq:var_decomp_one_term}
\end{equation}
If, for some sample having a non-zero contribution $c_r(P)$, the number of particles attaining this point $P$ goes to zero, then the generated variance will go to infinity. In deep-penetration problems, when playing analog games, very few particles will be able to reach the region surrounding the detector. The flights and collisions occurring in that region will most likely have a non-negligible relative contribution (as they are close to the detector), and thus a very small sampling intensity: $I_s(P)= n(P)/(c_r(P)N)\ll 1 $. This means that the variance term associated with these sampling events (\cref{eq:var_decomp_one_term}) will be extremely large, yielding an unacceptable variance. Actually, in such cases, some simulations simply cannot converge without using variance-reduction techniques.

We need then to avoid very small sampling intensities across phase space: in regions where the relative contribution is not negligible, we need to send a `sufficient' amount of particles. A simple way to enforce this condition is to use splitting and roulette to adapt the number of particles in phase space, so that
\begin{align}
    &I_s(P) \approx 1 \\
    \Leftrightarrow & \frac{n(P)}{N} \approx c_r(P)
\label{eq:constant_sampling_intensity}
\end{align}
across phase space. This population-control strategy is commonly known as `weight windows' \citep{booth_genesis_2006}.

A {\em target} weight $w_t(P)$ is first defined as a function of phase space. Then, a comparison is run between the particle weight and the target weight. If the weight of the particle is too large compared to the target weight, then splitting is applied to produce offspring with a weight closer to the target. Conversely, if the weight is too small, then roulette is applied to either kill the particle or increase its weight. Weight windows are a useful tool, since they ensure that the particle population at a point $P$ has a weight roughly equal to $w_t(P)$.

If we suppose that the application of weight windows yields $\mathbb{E}(w(P))\approx w_t(P)$, then requiring \cref{eq:constant_sampling_intensity} is equivalent to ensuring
\begin{equation}
    w_t(P) \approx \frac{\esp{R}}{M_1(P)} \propto M_1(P)^{-1},
    \label{eq:weight_target_simple}
\end{equation}
where we expressed $\espw{w(P)}$ as $\rho(P)N/n(P)$,
replaced $c_r$ in \cref{eq:constant_sampling_intensity} by its definition \cref{eq:def_rel_contrib} and simplified. The idea of imposing a target weight for weight windows inversely proportional to the importance $M_1(P)$, as suggested by \cref{eq:weight_target_simple}, is well established in the literature of variance-reduction methods \citep{booth_genesis_2006}. In practice, $M_1(P)$ is most often approximated using $I(\mathcal{P})$ a discretized version of $\chi^\dagger(\mathcal{P})$ obtained with a deterministic solver to apply the weight window at the emission point. If one wishes to apply weight window at the collision point, one can use 
\begin{equation}
    J(\mathcal{P}):=I(\mathcal{P})-\frac{{\nabla} I(\mathcal{P})\cdot \mathbf{\Omega}}{\Sigma_t(\mathcal{P})}
    \label{eq:approx_col_importance}
\end{equation} which corresponds to an approximation of $\psi^\dagger$\footnote{If the function $I$ converges towards $\chi^\dagger$, and ${\nabla} I$ converges towards ${\nabla} \chi^\dagger$ then $J$ converges towards $\psi^\dagger$.} . When using a discretized solution of the adjoint problem, the variance-reduction technique takes the name of Consistent Adjoint-Driven Importance Sampling (CADIS) \citep{wagner_automated_1998}; heuristic functions can also be successfully used, at the expense of some extensive fine-tuning in order to obtain satisfactory results  \citep[see for example the INIPOND method used in the Monte Carlo code TRIPOLI-4\textsuperscript{\textregistered},][]{hugot_overview_2024}.

Properly setting the multiplicative constant in \cref{eq:weight_target_simple} is crucial for the efficient application of the weight windows. If the constant is not adapted, it might trigger massive rouletting or splitting at the start of the particle life (i.e., after its birth is sampled). On the one hand, if the source is localized in phase space, the constant can be set so that the target weight is around unity in the vicinity of the source: particles born with unit weight avoid roulette and splitting at birth. When the CADIS method is applied, importance sampling is enforced also on the source, using
\begin{equation}
    \hat{S}(\mathcal{P}) = \frac{S(\mathcal{P})I(\mathcal{P})}{\int S(\mathcal{P}')I(\mathcal{P'})d\mathcal{P}'},
\end{equation}
where the integral is numerically evaluated. A particle will be thus created at point $\mathcal{P}$ with a weight
\begin{equation}
    w_\text{source}(\mathcal{P}) =\frac{\int S(\mathcal{P}')I(\mathcal{P'})d\mathcal{P}'}{I(\mathcal{P})} \approx \frac{\esp{R}}{\chi^\dagger(\mathcal{P})}
\end{equation}
corresponding exactly to the target weight calculated in \cref{eq:constant_sampling_intensity}, whence the `consistent' qualification for the CADIS method\footnote{Observe however that in the standard application of CADIS to weight windows only the source is sampled according to importance: flights and collisions are not modified.}.

\subsubsection{FoM-optimal strategy}
\label{sec:fomtimal_solutions}

The previous strategy was derived based on the following idea: ensuring a sampling intensity that is not too small is a necessary requirement to prevent the occurrence of excessive variances. For this purpose, a target value $I_s=1$ was proposed. This approach, however, might not be optimal for the FoM.

The FoM obeys \cref{eq:FoM}, which is rewritten here for convenience:
\begin{equation*}
    (\text{FoM})^{-1} \propto \left[ \int_\Pi c_r(P)I_s(P)dP\right]\left[ \int_\Pi c_r(P)\frac{1+\relvarw{w(P)}}{I_s(P)}\relvar{X_P}dP \right].
\end{equation*}
We seek then the function $I_s(P)$ that maximizes the FoM, i.e.~that minimizes the quantity $(\text{FoM})^{-1}$. We will assume that we can modify the sampling intensity $I_s(P)$ at will, by using some population-control technique. Furthermore, we will suppose that this process will not generate any variance by itself. This is in general false, as roulette involves a sampling procedure that needs to be accounted for in the variance-decomposition sum (see \cref{eq:decomp_var}). However, we will later show that in most cases this added variance is negligible compared to the variance of the other processes in the Monte Carlo game.

We will make the hypothesis that the weight variance $\relvarw{w(P)}$ is null, since all the weights are roughly equal to the target weight $w_t(P)$. Having a non-zero weight variance can only be detrimental to the FoM, as it affects negatively the variance (the variance part in \cref{eq:FoM} is a sum of positive or null terms that can only grow if $\relvarw{w(P)}$ is non-zero), while not improving the simulation time (this term does not occur in the time part of \cref{eq:FoM}). The relative contributions are fixed, since they depend only on the source, the response and the physical problem at hand. We will further assume that the sampling processes are fixed, i.e.~we do not apply any kind of importance sampling (and thus the zero variance is unreachable).

Let us first introduce the unitary relative variance of the result:
\begin{equation}
    \urelvar{R} = N\relvar{R},
\end{equation}
i.e.~the variance of the result for one source particle. We will also need the total sampling intensity:
\begin{equation}
    I_{s,tot} = \int_\Pi c_r(P)I_s(P)dP,
\end{equation}
which is an ersatz for the average simulation time required for one particle.

Achieving the minimum $(\text{FoM})^{-1}$ implies that the functional derivative of the $(\text{FoM})^{-1}$ with respect to $I_s(P)$ is null, namely
\begin{align}
    &\frac{\delta (\text{FoM})^{-1}}{\delta I_s(P)}= 0\nonumber\\
    \Leftrightarrow \qquad& c_r(P)\urelvar{R} - I_{s,tot} c_r(P) \frac{\relvar{X_P}}{I_s(P)^2} = 0\nonumber\\
    \Leftrightarrow \qquad& I_s(P) = \sqrt{\frac{I_{s,tot}}{\urelvar{R}}\relvar{X_P}}.
    \label{eq:exact_optimal_sampling_intensity}
\end{align}
The presence of the terms $I_{s,tot}$ and $\urelvar{R}$ in \cref{eq:exact_optimal_sampling_intensity} makes it difficult to interpret the formula. However, since they do not depend on the considered point $P$, we can state that the optimal value is 
\begin{equation}
    I_s(P) \propto \sqrt{\relvar{X_P}}.
    \label{eq:optimal_sampling_intensity}
\end{equation}
The multiplicative constant has no influence on the FoM: setting $\hat{I}_s(P)=\lambda I_s(P)$ and inserting it into \cref{eq:FoM} shows that the $\lambda$ from the time component and the $\lambda$ from the variance component cancel. This is compatible with \cref{eq:exact_optimal_sampling_intensity}, since the term $I_{s,tot}/\urelvar{R}$ is proportional to $\lambda^2$.

The result obtained in \cref{eq:optimal_sampling_intensity} suggests that we should spend more computing power sampling the processes having a larger intrinsic variance, which decreases their impact on the final variance. This can be taken into account e.g.~in the target weight of the weight windows by setting:
\begin{equation}
    w_t(P) \propto \left(M_1(P) \sqrt{\relvar{X_P}} \right)^{-1}.
    \label{eq:weight_target_precise}
\end{equation}

This formula, which provides an improved version of \cref{eq:weight_target_simple}, comes however with several caveats. First, obtaining the term $\relvar{X_P}$ is usually harder than obtaining $M_1(P)$, which can be either obtained via a deterministic adjoint solver, or estimated by simple heuristics. We will nonetheless show that a very crude approximation can be obtained in some situations (see \cref{sec:example_pure_el_scat}). The correction applied by taking into account $\relvar{X_P}$ is usually not as important as the $M_1(P)^{-1}$ shape. This can be understood as follows: in problems where a weight window is required, the importance usually varies over several orders of magnitude between the source and the detector, whereas the intrinsic variance usually displays milder variations. Finally, and most important, the intrinsic variance $\relvar{X_P}$ may vary between a sampling and the next. For instance, consider a problem where sampling different outgoing energies might lead to roughly the same importance, leading to a small intrinsic variance for this process, whereas the next flight determining whether the particle will hit the detector zone might have a huge intrinsic variance. This high heterogeneity means that strictly following \cref{eq:weight_target_precise} will lead to applying weight correction roughly at each sampling process. Remember that we derived this ideal sampling intensity under the hypothesis that the variance added by the population control was negligible. If one is able to evaluate exactly \cref{eq:weight_target_precise} and applies weight control before each sub-samplings, the variance added by the population-control procedure will not be negligible\footnote{The derivation of exactly how much variance is added during a population control process will be provided in \cref{sec:pop_control}.} and will clearly worsen the result compared to the simpler version presented in \cref{eq:weight_target_simple}. However, \Cref{eq:weight_target_precise} can still be useful if applied carefully, as shown in \cref{sec:example_pure_el_scat}.

\subsection{Variance induced by population control}
\label{sec:pop_control}

Population-control techniques such as roulette or splitting involve a random process (for splitting this is only true if the weight before splitting is not a multiple of the target weight). This random process will generate some intrinsic variance that needs to be taken into account in the variance-decomposition formula \cref{eq:decomp_var}. 

As the random process describing population control are weight-dependent, we distinguish the different samplings for the different weights (as explained in \cref{sec:context_and_def}). We will denote $P_w$ and $X_{P_w}$ the entering point and the corresponding sampling specific to weight $w$ in the following section. Correspondingly, we will refer to $P$ as corresponding to all $P_w$ irrespective of the weights. As having different points for different weights renders the use of the variance decomposition formula \cref{eq:decomp_var} unwieldy, we will derive a `condensed' formula that integrates all the variance contributions from all points $P_w$ for $w\in[0,\infty[$ at a given point $P$.

Let us first study the formula for weight control on a single particle, including splitting and roulette. A fully general weight-control process might have a threshold weight for rouletting and a specific target weight $w_t$ for particles exiting roulette, and it might include splitting or not. For the sake of generality, we will assume that our process aims at generating offsprings (zero, one or several) at a target weight $w_t(w)$ depending on the entering weight $w$. This encompasses roulette (where the number of offsprings is either zero or one) and splitting (the number of offsprings being one or higher). To keep the description simple, we introduce $f:= w/w_t(w)$. The weight-control process is described as follows: a particle of weight $w$ is turned into $\lfloor f \rfloor$ offsprings of weight $w_t(w)=f(w)w$ with probability $\lfloor f+1\rfloor -f$, and into $\lfloor f +1 \rfloor$ particles of weight $w_t(t)$ with probability $f-\lfloor f \rfloor$.

By integrating the different contributions to the variance of the different samplings at the different weights $w$ in $dP$ around point $P$, we show in \cref{sec:proof_split_roul} that the total contribution of the variance in $dP$ around $P$ is:  

\begin{equation}
    \var{\text{Pop. cont. in }dP\text{ around } P} = \frac{c_r(P)dP}{I_s(P)} \int_{w=0}^\infty \frac{n(P,w)}{n(P)}\frac{(w_t(w))^2}{\espw{w(P)}^2} (f-\lfloor f \rfloor)(\lfloor f+1\rfloor-f) dw,
    \label{eq:var_pop_control}
\end{equation}
where $n(w,P)dPdw$ is the number of particles in $dP$ around $P$, with weight $dw$ around $w$.

The formula in \cref{eq:var_pop_control} can be understood as following. First, the terms $c_r(P)$ and $I_s(P)$ are the standard relative contribution and sampling intensity terms, respectively, emerging in all the sampling processes. The relative variance of the weights does not occur. The term within the integral is the equivalent of the intrinsic variance for the whole weight-control process. The integral acts on the weight distribution $n(P,w)/n(P)$. Recall that $f=w/w_t(w)$ corresponds to how many offsprings we want to create ideally, and depends on $w$: the factor $(f-\lfloor f \rfloor)(\lfloor f+1\rfloor-f)$ is null if $f$ is an integer (no stochastic process is involved, and no variance is added), and maximal if $f$ is a half-integer (with a value of $1/4$). Finally, the last term $(w_t(w))^2/\espw{w(P)}^2$ amplifies the variance if we are mostly doing roulette ($w_t>\espw{w}$), and reduces it if we are mostly doing splitting ($w_t<\espw{w}$). For weight windows, the term $w_t$ is usually close to the center of the window (it can even be set to be constant and precisely equal to the center of the window). In the particular case where we only use weight control to eliminate the weight variance, i.e. if we want to align the whole weight distribution with its average value $\espw{w(P)}$, then this factor is equal to one. Since the function involving $f$ has a maximum of $1/4$, in this case the intrinsic variance term cannot exceed $1/4$. This provides a rough estimate of the variance added by eliminating the variance of the weights over the whole population by sending all the weight to their average $w_t(w)=\espw{w(P)}$.

\subsection{Importance sampling and path stretching}

Importance sampling refers to a large class of variance reduction techniques whose common feature is to modify the sampling laws of the different samples $X_P$ proportionally to an arbitrary function $\mathcal{I}(P)$, that is user-defined and aims at approximating $M_1(P)$ either through a deterministic calculation or through some heuristic. $\mathcal{I}$ is usually called `importance', hence the `importance samplings'. To avoid confusion with $M_1$ we will refer to it as `approximated importance' or `user importance'. In the field of transport, we will denote $I(\mathcal{P})$ the `approximated emission importance' and $J(\mathcal{P})$ the `approximated collision importance'. Note that $J$ can be computed from $I$ using \cref{eq:approx_col_importance}.

For example, consider the sampling of a flight from point $\mathcal{P}$ where we want to sample the next collision point $\mathcal{P}'$. Instead of sampling from the analogue distribution 
\begin{align}
    T(\mathcal{P}\rightarrow \mathcal{P}') =&\Sigma_t(\mathbf{r}',E') \exp\left(-\int_0 ^{|\mathbf{r}-\mathbf{r}'|}  \Sigma_t(\mathbf{r}+l\mathbf{\Omega},E)dl\right) \nonumber \\
    &\times\frac{\delta\left(\frac{\mathbf{r}'-\mathbf{r}}{|\mathbf{r}-\mathbf{r}'|} - \mathbf{\Omega} \right)}{|\mathbf{r}-\mathbf{r}'|^2} \delta \left(E-E' \right)  \delta(\mathbf{\Omega}-\mathbf{\Omega}'),
    \label{eq:decomposed_direct_T}
\end{align}
 we sample from the modified distribution:
\begin{equation}
    \hat{T}(\mathcal{P}\to \mathcal{P}') \propto T(\mathcal{P}\to \mathcal{P}') J(\mathcal{P}'),
    \label{eq:transport_ker_importance_sampling}
\end{equation}
and apply a weight correction
\begin{equation}
    \hat{m}=\frac{T(\mathcal{P}\to \mathcal{P'})}{\hat{T}(\mathcal{P}\to \mathcal{P}')},
    \label{eq:transport_ker_importance_sampling_weight}
\end{equation}
to keep the sampling unbiased.

Importance sampling works similarly to zero-variance by reducing the intrinsic variance of the samplings $\relvar{X_P}$. Zero-variance could even be considered as the limit of importance sampling when the approximated importance $\mathcal{I}(P)$ converges towards the actual importance $M_1(P)$. Note that zero-variance procedures need to be applied to all samplings in the Monte Carlo game, including samplings that might create multiple outgoing particles. The actual general equation for zero-variance will be provided in Sec.~\ref{sec:zero_variance} (see \cref{eq:zv_general_sample}), and it involves the importance of $P'_k$, the number of offspring $K$ and the weight multipliers $m_k$.

Let us illustrate the reduction of the intrinsic variance using our previous example of flight. The importance of the sampling of the next collision point $\mathcal{P}'$ using $T(\mathcal{P}\to \mathcal{P'})$ was
\begin{equation}
    M_1(X_P)= \psi^\dagger(\mathcal{P'}),
    \label{eq:importance_sampling_ref_importance}
\end{equation}
using importance sampling according to \cref{eq:transport_ker_importance_sampling,eq:transport_ker_importance_sampling_weight} the importance of the sampling is now
\begin{equation}
    M_1(\hat{X}_P)=\frac{T(\mathcal{P}\to \mathcal{P'})}{\hat{T}(\mathcal{P}\to \mathcal{P}')}\psi^\dagger(\mathcal{P'}) \propto \frac{\psi^\dagger(\mathcal{P}')}{J(\mathcal{P'})}.
    \label{eq:importance_sampling_importance}
\end{equation}
As $\relvar{\hat{X}_P}=\relvar{M_1(\hat{X}_P)}$, the closest the approximated function $J$ can be to the actual collision importance $\psi^\dagger$, the more constant will $M_1(\hat{X}_P)$ be and the smaller will $\relvar{M_1(\hat{X}_P)}$ be. In the limit where we use the actual importance $\psi^\dagger$ as $J$, $M_1(\hat{X}_P)$ is constant and this sampling is `zero-variance' (it adds no variance to the variance of the result $R$). Note that if an inappropriate approximated importance $J$ is used, the variable $M_1(\hat{X}_P)$ could have bigger variations than the original $M_1(X_P)$ (from \cref{eq:importance_sampling_ref_importance}) and the intrinsic variance could be worsened.

Another benefit of importance sampling (provided $\mathcal{I}$ is close enough to $M_1$) is that it naturally `pushes' the particles towards important regions, by avoiding premature captures before the detector is reached, or by favoring flights that point towards the detector. In addition, importance sampling will naturally lead to reducing under-sampling (i.e. avoiding regions of very low sampling intensity $I_s$). However, this can backfire if no capture occurs and particles have fewer opportunities to die, leading to longer calculations, especially if the sampling intensity is already boosted by a population-control scheme.

Path stretching is a form of importance sampling applied to the flight kernel only. Sampling a non-analog flight kernel is usually easier than sampling a non-analog collision kernel, as it does not involve tampering with the wide variety of sampling procedures included in modern nuclear data libraries  \citep[see, for example, the sampling procedures of ACE files produced by the NJOY nuclear data processing code,][]{conlin_compact_2019}. To use path stretching, the user simply samples $\hat{T}$ similarly to the analogue method (see \cref{eq:decomposed_direct_T}) but using a modified total cross-section
\begin{equation}
    \hat{\Sigma}_t = \Sigma_t - \mathbf{\Omega}\cdot \frac{{\nabla} I}{I}.
\end{equation}
Doing so effectively samples from the modified distribution:
\begin{align}
    \hat{T}(\mathcal{P}\to \mathcal{P}') =& \hat{\Sigma}_t(\mathbf{r}',E',\mathbf{\Omega}) \exp\left(-\int_0 ^{|\mathbf{r}-\mathbf{r}'|}  \hat{\Sigma}_t(\mathbf{r}+l\mathbf{\Omega},E,\mathbf{\Omega})dl\right) \nonumber \\
    &\times\frac{\delta\left(\frac{\mathbf{r}'-\mathbf{r}}{|\mathbf{r}-\mathbf{r}'|} - \mathbf{\Omega} \right)}{|\mathbf{r}-\mathbf{r}'|^2} \delta \left(E-E' \right)  \delta(\mathbf{\Omega}-\mathbf{\Omega}')\nonumber \\
    =& \frac{J(\mathcal{P'})}{I(\mathcal{P}')}\Sigma_t(\mathbf{r}',E') \exp\left(-\int_0 ^{|\mathbf{r}-\mathbf{r}'|}  \Sigma_t(\mathbf{r}+l\mathbf{\Omega},E)dl\right)\nonumber\\&\times\exp\left(\int_0 ^{|\mathbf{r}-\mathbf{r}'|}  \frac{\mathbf{\Omega}\cdot {\nabla}I(\mathbf{r}+l\mathbf{\Omega},E,\mathbf{\Omega})}{I(\mathbf{r}+l\mathbf{\Omega},E,\mathbf{\Omega})}dl\right) \nonumber \\
    &\times\frac{\delta\left(\frac{\mathbf{r}'-\mathbf{r}}{|\mathbf{r}-\mathbf{r}'|} - \mathbf{\Omega} \right)}{|\mathbf{r}-\mathbf{r}'|^2} \delta \left(E-E' \right)  \delta(\mathbf{\Omega}-\mathbf{\Omega}')\nonumber \\
    =& \frac{J(\mathcal{P'})}{I(\mathcal{P}')}T(\mathcal{P}\to \mathcal{P}')\exp\left(\left[\ln(I(\mathcal{\tilde{P}})) \right]^\mathcal{P'}_\mathcal{P}\right) \nonumber \\
    =& T(\mathcal{P}\to \mathcal{P}')\frac{J(\mathcal{P'})}{I(\mathcal{P})},
    \label{eq:flight_ker_path_stretching}
\end{align}
where $J$ is computed from $I$ using \cref{eq:approx_col_importance}. One can then apply the weight correction $I(\mathcal{P})/J(\mathcal{P'})$ to preserve unbiasedness.

We can see that \cref{eq:flight_ker_path_stretching} corresponds to \cref{eq:transport_ker_importance_sampling}, proving that path stretching is indeed a form of importance sampling. When $I$ converges towards $\chi^\dagger$, and $\nabla I$ converges toward $\nabla \chi^\dagger$, then $J$ converges toward $\psi^\dagger$, and the variance-reducing properties of importance sampling discussed in this section will apply to path stretching.

\section{Application to zero-variance games}
\label{sec:zero_variance}

In this section we will use the variance-decomposition formula in \cref{eq:decomp_var} to obtain the general rules of the zero-variance Monte Carlo games. It will then be applied to both direct and adjoint transport problems.

We will show that any game that falls within the framework of this paper (as defined in \cref{sec:context_and_def}) admits a `modified game' having the same generalized phase space but with the law of the samplings $X_P$ modified, that yields the same expected response $\esp{R}$ with a null variance. Note that, even if most of this work could be generalized without the hypothesis of positivity of the Monte Carlo game (positivity of the weight multipliers $m_k$ and of the estimator function $h$), positivity is a key element absolutely required for a zero-variance game to exist.

We set out to obtain the rules required for the different sampling events of the Monte Carlo game to lead to zero variance on the final result. First, note that 
the integrand in the integral over phase space $\Pi$ in the variance-decomposition formula (\cref{eq:decomp_var}) are positive. This means that the integrand needs to be zero for all $P$ in order to attain zero variance.

At a given point $P$, the integrand reads:
\begin{equation}
c_r(P)\frac{1+\relvarw{w(P)}}{I_s(P)}\relvar{X_P}.
\end{equation}
The term $1+\relvarw{w(P)}$ cannot be zero, since $\relvarw{w(P)}$ is positive or null. The term $1/I_s(P)$, being the ratio of the relative contribution of $P$ to the fraction of particles reaching $P$, can only vanish if the relative contribution $c_r(P)$ is zero. This leaves two options for the integrand to be zero: either the relative contribution $c_r(P)$ of the sampling is zero (which means that it has no influence on the result $R$), or the intrinsic variance $\relvar{X_P}$ itself is zero. 

Ensuring a vanishing intrinsic variance means that the random variable $M_1(X_P)$ is deterministic:
\begin{align}
    &\relvar{X_P} = \relvar{M_1(X_P)} = 0\nonumber \\
    \Leftrightarrow& M_1(X_P)\text{ is deterministic}\nonumber\\
    \Leftrightarrow& \sum_{k=1}^K m_k M_1(P'_k) = \text{constant}.
    \label{eq:zv_implication}
\end{align}
\Cref{eq:zv_implication} allows deducing the following implications for all the different types of sampling events needed in a standard direct Monte Carlo game.

The general idea is the following. Assume that the original distribution of $X_P$ in the game is $f(x)=\mathbb{P}(X_P = x)$, with $x$ being any of the possible outcomes of $X_P$. Keeping the same generalized phase-space, we only allow ourselves to modify the probability of the different outcomes of $X_P$, sampling instead from a modified distribution $\hat{f}(x)$. The requirement of unbiasedness implies that the new weight multipliers $\hat{m}_k$ satisfy
\begin{equation}
    \hat{m}_k= m_k \frac{f(x)}{\hat{f}(x)}.
\end{equation}
Applying the zero-variance condition \cref{eq:zv_implication} to this modified game leads to:
\begin{align}
    &\sum_{k=1}^K \hat{m}_k M_1(P'_k) = \text{constant}\nonumber\\
    \Leftrightarrow& \frac{f(x)}{\hat{f}(x)} \sum_{k=1}^K m_k M_1(P'_k) = \text{constant}\nonumber\\
    \Leftrightarrow& \hat{f}(x) \propto f(x) \times \sum_{k=1}^K m_k M_1(P'_k).
    \label{eq:zv_general_sample}
\end{align}
From \cref{eq:zv_general_sample} we deduce that the zero-variance game is obtained by correcting the probability of sampling each outcome $x$ of $X_P$ proportionally to its importance.

If any of the outcomes $x$ are not sampled using the importance-weighted $\hat{f}$, then the weight multipliers will be different, and \cref{eq:zv_implication} will not be satisfied. Once the random events $X_P$ composing the Monte Carlo game are fixed, there is a unique way to sample them to obtain zero variance, except for random events that lead to no contributions because they never occur and/or have no importance. If the game is modified in other ways, for example by changing the estimator $h$, by adding/removing spitting or roulette at different points in phase space, or by switching from branching to branchless collision sampling, then there exist other zero-variance sampling strategies for this modified game\footnote{In transport problems, this implies that, for a given geometry, source and detector response function $\eta_\psi$, there exist many zero-variance games yielding the sought response $\esp{R}$, since there exist many partially-unbiased estimators, and the game can be played using branching or branchless histories \citep[see e.g.][]{lux_monte_2018}.}.

We will now detail how this applies to actual sampling in a direct transport Monte Carlo game.

\subsection{Zero-variance games for direct transport}
\label{sec:zv_direct}
In this section we will modify the sampling laws of an analog\footnote{An `analog' game could be described as a simulation where the sampling process follows as closely as possible the rules of underlying physical process. In practice, not only this is unfeasible due to the limited amount of information provided by nuclear data libraries, but also somewhat useless, since the Boltzmann equation (whose solution the game is attempting at estimating in the first place) is only concerned with the average behavior of the particle population. For this reason, we refer to the less restrictive definition of `analog' games as given in \citet{lux_monte_2018}.} game to obtain a zero-variance game. This will allow us to recover the known formulas for the zero-variance source, flight and collision kernels derived from the moment equations in Chap. V, Part VIII of \citet{lux_monte_2018}.

\subsubsection{Source}

The first event that occurs in a Monte Carlo game is the sampling of the source $S$. This sampling occurs for a single particle, so the sum over $k$ reduces to just one term. Once the source has been sampled, the particle is in an emitted state. Therefore, $M_1(P')$ corresponds to the importance $\chi^\dagger$ of a particle starting a flight at point $P'$. We thus have:
\begin{equation}
    M_1(P')= \chi^\dagger(\mathcal{P}').
\end{equation}

If we sample from a modified importance-weighted distribution $\hat{S}$ instead of the natural source density $S$, unbiasedness is preserved by setting the non-analog weight factor $\hat{m}$ to:
\begin{equation}
    \hat{m}(P') = \frac{S(\mathcal{P}')}{\hat{S}(\mathcal{P}')}.
\end{equation}
\Cref{eq:zv_implication} then becomes:
\begin{align*}
    & \hat{m}(P') \chi^\dagger(\mathcal{P}') = \text{constant}\\
    \Leftrightarrow &  \frac{S(\mathcal{P}')}{\hat{S}(\mathcal{P}')}\chi^\dagger(\mathcal{P}')= \text{constant}\\
    \Leftrightarrow & \hat{S}(\mathcal{P}') \propto S(\mathcal{P}')\chi^\dagger(\mathcal{P}).
\end{align*}
Since the distribution $\hat{S}$ must be normalized, we impose
\begin{equation}
    \hat{S}(\mathcal{P}') = \frac{S(\mathcal{P}')\chi^\dagger(\mathcal{P}')}{\int S(\mathcal{P})\chi^\dagger(\mathcal{P})d\mathcal{P}}.
    \label{eq:zv_source}
\end{equation}

\subsubsection{Flights}
\label{sec:zv_direct_flight}

Flight events also occur for a single particle (we suppose here that no splitting occurs during the flight), and $K=1$. At the end of the flight the particle is in a state of entering collision so 
\begin{equation}
   M_1(P')=\psi^\dagger(\mathcal{P'}).
   \label{eq:importance_for_flight_col_est}
\end{equation}
Here, note that \cref{eq:importance_for_flight_col_est} is true as we use a collision estimator. When using other estimators (and in particular track length estimators) \cref{eq:importance_for_flight_col_est,eq:zv_flight_ker} are not true. This will be treated in detail in \cref{sec:non_col_estimators}.

Using \cref{eq:zv_general_sample}, we finally obtain:
\begin{equation}
    \hat{T}(\mathcal{P}\to \mathcal{P}') \propto T(\mathcal{P}\to \mathcal{P}') \psi^\dagger(\mathcal{P}'),
    \label{eq:zv_flight_ker}
\end{equation}
which is a classical formula for zero-variance for collision estimators \citep[see e.g. ][]{lux_monte_2018}).

\subsubsection{Collisions}
\label{sec:zv_direct_collision}

A peculiar feature of the collision event is that it usually consists of several steps: sampling the nuclide, then the reaction channel, sometimes the multiplicity, and finally the outgoing energy and direction for the secondary particles.

Without loss of generality, we will encode the non-analog sampling of nuclides (indexed by $i$) and reaction channels (indexed by $j$) by introducing non-analog microscopic cross-sections denoted $\hat{\sigma}$, and assume that we sample the nuclides and reaction channels using these cross-sections:
\begin{equation}
    \hat{\mathbb{P}}(\{i,j\})= \frac{N_i\hat{\sigma}_{i,j}}{\hat{\Sigma}_t},
\end{equation}
where $\hat{\Sigma}_t$ is the total macroscopic cross-section associated to $\hat{\sigma}_{i,j}$.
In full generality, the $\hat{\sigma}$ can also depend on position $\mathbf{r}$ and direction $\mathbf{\Omega}$, in addition to energy $E$. The weight correction applied after choosing nuclide and reaction channel will be:
\begin{equation}
    \hat{m} = \frac{\sigma_{i,j}(\mathcal{P})}{\Sigma_t(\mathcal{P})} \times \frac{\hat{\Sigma}_t(\mathcal{P})}{\hat{\sigma}_{i,j}(\mathcal{P})}.
\end{equation}
The importance $M_1$ of a collision event involving the pair $\{i,j\}$ (denoted $\psi^\dagger_{i,j}$ in the following) can be evaluated by using:
\begin{equation}
    \psi^\dagger_{i,j} (\mathbf{r},E,\mathbf{\Omega}) =  \iint \nu_{i,j}(E)f_{i,j}(E\to E', \mathbf{\Omega}\cdot \mathbf{\Omega'})\chi^\dagger(\mathbf{r},E',\mathbf{\Omega'})d\mathbf{\Omega'}dE',
    \label{eq:importance_channel_ij}
\end{equation}
where $\nu_{i,j}$ is the average multiplicity of the reaction channel $\{i,j\}$ and $f_{i,j}$ is the outgoing distribution law of channel $\{i,j\}$. Applying \cref{eq:zv_general_sample} we get
\begin{align}
     \hat{\mathbb{P}}(\{i,j\}) &\propto \frac{N_i\sigma_{i,j}}{\Sigma_t}\times \psi^\dagger_{i,j}\nonumber\\
    \Leftrightarrow\qquad\qquad \hat{\sigma}_{i,j} &\propto \sigma_{i,j}\psi^\dagger_{i,j} 
    \label{eq:zv_choice_channel}
\end{align}

The multiplicity can be handled in several ways. If one chooses to only apply a weight correction $\nu_{i,j}$ on the particle, no random procedure is actually taking place, so no variance is added here: this procedure will result in a branchless zero-variance game. If, however, one chooses to create a number of offsprings corresponding to the value $\nu_{i,j}$, then a sampling will inevitably occur when $\nu_{i,j}$ is not integer. In full generality, let us denote $\mathbb{P}(K=n)$ the probability of sampling $n$ offsprings in the analog game\footnote{As mentioned in the introduction of this section, the notion of `analog' game is not well-defined. Here, the formalism employed for multiplicity covers both the use of the exact distribution of multiplicities or the use of average multiplicities $\nu_{i,j}$, in which case we sample either $\lfloor\nu_{i,j}\rfloor$ or $\lfloor\nu_{i,j}+1\rfloor$ outgoing particles}, and $\hat{\mathbb{P}}(K=n)$ in the non-analog game. The weight correction is then $\hat{m} = \mathbb{P}(K=n) / \hat{\mathbb{P}}(K=n)$. For the sake of simplicity, let us suppose that all the offsprings have the same outgoing distribution: this means that each offspring has the same importance, denoted $M_1(\text{offspring})$. The importance of sampling $n$ offsprings is then $n$ times the importance of one offspring. \Cref{eq:zv_implication} means that
\begin{equation}
    \sum_{k=1}^{K=n} \hat{m} M_1(\text{offspring}) =  n \frac{\mathbb{P}(K=n)}{\hat{\mathbb{P}}(K=n)} M_1(\text{offspring})=\text{constant},
\end{equation}
for all possible numbers $n$ of offsprings. This leads to
\begin{equation}
    \hat{\mathbb{P}}(K=n) \propto n \mathbb{P}(K=n).
\end{equation}

The last step is to sample the outgoing distribution law $f_{i,j}$. Again, if we sample from $\hat{f}_{i,j}$, the weight correction is 
\begin{equation}
    \hat{m} = \frac{f_{i,j}(\mathcal{P}\to \mathcal{P}')}{\hat{f}_{i,j}(\mathcal{P}\to \mathcal{P}')},
\end{equation}
and the importance of the outgoing state $P'$ is simply the emission importance $\chi^\dagger(\mathcal{P}')$. Using \cref{eq:zv_general_sample} leads to:
\begin{equation}
    \hat{f}_{i,j}(\mathcal{P}\to \mathcal{P}') \propto f_{i,j}(\mathcal{P}\to \mathcal{P}') \chi^\dagger(\mathcal{P}').
\end{equation}

\subsubsection{Other procedures}

During the zero-variance game, one might be tempted to apply other procedures, such as splitting and roulette. As previously discussed while treating multiplicity (cf.~\cref{sec:zv_direct_collision}), splitting a particle into an integer number of offsprings can be done deterministically, i.e.~without sampling. Therefore, this adds no variance, and in particular does not alter the zero-variance nature of the simulation.

Applying Russian roulette, on the contrary, involves a sampling to determine whether the particle survives or dies. Denoting $p_s$ the probability of surviving, we then have a weight multiplier $m=1/p_s$ to ensure unbiasedness if the particle survives. Obtaining the zero-variance non-analog probability corresponding to this process yields $\hat{p}_s=1$: death has an importance of zero, so the only option possible is to sample survival. This leads to a non-analog weight multiplier $\hat{m} = p_s m = 1$, which is equivalent to taking no action. Therefore, there is no zero-variance roulette procedure.

\subsection{Zero-variance games for adjoint transport}
\label{sec:zv_adjoint}

The variance-decomposition formula can be also applied to adjoint Monte Carlo games. In this context, the function $M_1$ describes the importance of adjoint particles with respect to the adjoint detector. The bi-adjoint of an operator is the operator itself. Thus, the importance function for an adjoint game is obtained by applying the regular (direct) transport operator to the source, which corresponds to the detector response function of the adjoint problem. The importance functions in the adjoint game will be then provided by the collision and emission densities obtained with a source density corresponding to the response function of the adjoint detector.

\subsubsection{Source}

Let us denote $S^\dagger$ the source density of adjunctons, and $\hat{S}^\dagger$ the modified distribution from which particles are sampled. The weight multiplication associated with a particle sampled at point $P'$ is then:
\begin{equation}
    \hat{m} = \frac{S^\dagger(\mathcal{P'})}{\hat{S}^\dagger(\mathcal{P'})}.
    \label{eq:adj_source_weight_corr}
\end{equation}
If the adjoint problem being simulated follows the usual convention, and uses a collision-based response function $\eta_\psi$ as the adjoint source, the sampled adjunctons carry importance at collision, and their importance function $M_1(P')$ is the collision density $\psi(\mathcal{P'})$. Applying \cref{eq:zv_implication}, with $K=1$, $\hat{m}_k$ from \cref{eq:adj_source_weight_corr} we then get the zero-variance source distribution:
\begin{align}
    \frac{S^\dagger(\mathcal{P'})}{\hat{S}^\dagger(\mathcal{P'})}\psi(\mathcal{P'})&=\text{constant}\nonumber\\
    \Leftrightarrow \qquad\qquad\hat{S}^\dagger(\mathcal{P}') &\propto S^\dagger(\mathcal{P}')\psi(\mathcal{P}').
\end{align}

\subsubsection{Flights}

The adjoint flight kernel $T^\dagger(\mathcal{P}'\to \mathcal{P}) = T(\mathcal{P}\to \mathcal{P}')$ is, in general, not normalized in $\mathcal{P}$. One thus samples from the arbitrary normalized distribution $\hat{T}^\dagger(\mathcal{P}'\to \mathcal{P})$, and applies a weight correction\footnote{Note that in most papers the proposed distribution is a simple direct flight kernel $T(\mathcal{P}'\to \mathcal{P})$, which leads to a weight correction $\Sigma_t(\mathcal{P}')/\Sigma_t(\mathcal{P})$ \citep{irving_adjoint_1971,hoogenboom_adjoint_1977}. This is, in general, not the zero-variance distribution.}:
\begin{equation}
    \hat{m} = \frac{T^\dagger(\mathcal{P}'\to \mathcal{P})}{\hat{T}^\dagger(\mathcal{P}'\to \mathcal{P})}.
    \label{eq:weight_corr_adjoint_flight}
\end{equation}

Using a collision estimator (that scores at the end of the adjoint flight), the importance $M_1(P)$ of the adjoint particle flying from $P'$ to $P$ is:
\begin{equation}
    M_1(P) = \chi(\mathcal{P}).
    \label{eq:conv_flight_importance_adjoint}
\end{equation}
Applying \cref{eq:zv_implication} with $K=1$, we then get the zero-variance distribution:
\begin{equation}
    \hat{T}^\dagger(\mathcal{P}'\to \mathcal{P}) \propto T^\dagger(\mathcal{P}'\to \mathcal{P})  \chi(\mathcal{P}).
\end{equation}

\subsubsection{Collisions}

In the literature, several methods have been proposed to sample the adjoint collisions \citep{irving_adjoint_1971, hoogenboom_adjoint_1977,matteis_new_1978}. We will here follow the notation and conventions proposed in \citet{rovel_general_2025}.

The selection of the reaction channel $\{i,j\}$ for an adjoint collision is usually done by using arbitrary adjoint cross-sections $\sigma^\dagger_{i,j}$. This selection applies the weight multiplication factor
\begin{equation}
    \hat{m} = \frac{\Sigma_t^\dagger(\mathcal{P}')}{N_i(\mathcal{P}')\sigma^\dagger_{i,j}(\mathcal{P}')},
\end{equation}
where $N_i$ is the concentration of nuclide $i$.

After choosing the channel $\{i,j\}$, the adjoint particle is at the (generalized phase space) point where a direct particle would be after being emitted at point $P'$ by reaction $\{i,j\}$. This means that the importance $M_1$ is the emission density by $\{i,j\}$, denoted $\chi_{i,j}(\mathcal{P}')$:
\begin{equation}
    \chi_{i,j}(\mathcal{P}') = \int _{\mathbf{\Omega}} \int_{E} \psi(\mathbf{r},E,\mathbf{\Omega})\frac{N_i(\mathbf{r}) \sigma_{i,j}(E)}{\Sigma_t(\mathbf{r},E)}f_{i,j}(E\to E', \mathbf{\Omega} \cdot \mathbf{\Omega'})dE d\mathbf{\Omega}.
\end{equation}
Applying \cref{eq:zv_implication} then yields:
\begin{align}
    \frac{\Sigma_t^\dagger(\mathcal{P'})}{\sigma_{i,j}^\dagger (\mathcal{P}')N_i}\chi_{i,j}(\mathcal{P}')&= \text{constant}\nonumber\\
    \Leftrightarrow\qquad\qquad\qquad\sigma_{i,j}^\dagger (\mathcal{P}') &\propto \frac{\chi_{i,j}(\mathcal{P}')}{N_i(\mathbf{r})}.
\end{align}

After sampling the reaction channel, one needs to find the entering point $\mathcal{P}$ by sampling the arbitrary adjoint distribution law $f^\dagger_{i,j}(\mathcal{P}\to \mathcal{P}')$, and apply the weight multiplication
\begin{equation}
    \hat{m} = \frac{\nu_{i,j}(\mathcal{P})f_{i,j}(\mathcal{P}\to \mathcal{P}')}{f^\dagger_{i,j}(\mathcal{P}'\to \mathcal{P})},
\end{equation}
which corresponds to a correction factor that ensures the unbiasedness of the collision sampling \citep{rovel_general_2025}. Right after sampling the outgoing point $\mathcal{P}$, the adjoint particle will be in the same point as a direct particle would be after entering collision at point $\mathcal{P}$, and choosing nuclide $i$ and reaction $j$. The importance function $M_1$ is thus the collision density on $\{i,j\}$, denoted $\psi_{i,j}(\mathcal{P})$, namely
\begin{equation}
    \psi_{i,j}(\mathcal{P}) = \psi(\mathcal{P})\frac{N_i(\mathbf{r})\sigma_{i,j}(E)}{\Sigma_t(\mathbf{r},E)}.
\end{equation}
Applying \cref{eq:zv_implication} finally yields the following:
\begin{equation}
    f^\dagger_{i,j}(\mathcal{P}'\to \mathcal{P}) \propto \psi(\mathcal{P})\frac{N_i(\mathbf{r})\sigma_{i,j}(E)}{\Sigma_t(\mathbf{r},E)}\nu_{i,j}(E)f_{i,j}(\mathcal{P}\to \mathcal{P}').
\end{equation}

\section{Applications of the variance-decomposition formula}
\label{sec:examples}

In this section, we will check the validity of the formula of variance-decomposition by actually evaluating all its terms in a few relevant examples. These examples also serve as an illustration of how the different terms present in the formula behave in concrete examples to yield the total variance. 

Note that evaluating all the terms to compute the variance, as is done in this section, does not correspond to the expected use of the formula in real-world problems. Indeed, evaluating the formula requires solving both the direct and adjoint problem at hand. Even though actually evaluating the formula could prove useful on a simplified version of the problem or via deterministic methods, the expected use of the formula is more to serve as a new mathematical framework distinct from the moment equations.

\subsection{A discrete problem}
\label{sec:example_discrete_problem}

\begin{figure}[ht]
    \centering
    \includegraphics[width=\textwidth]{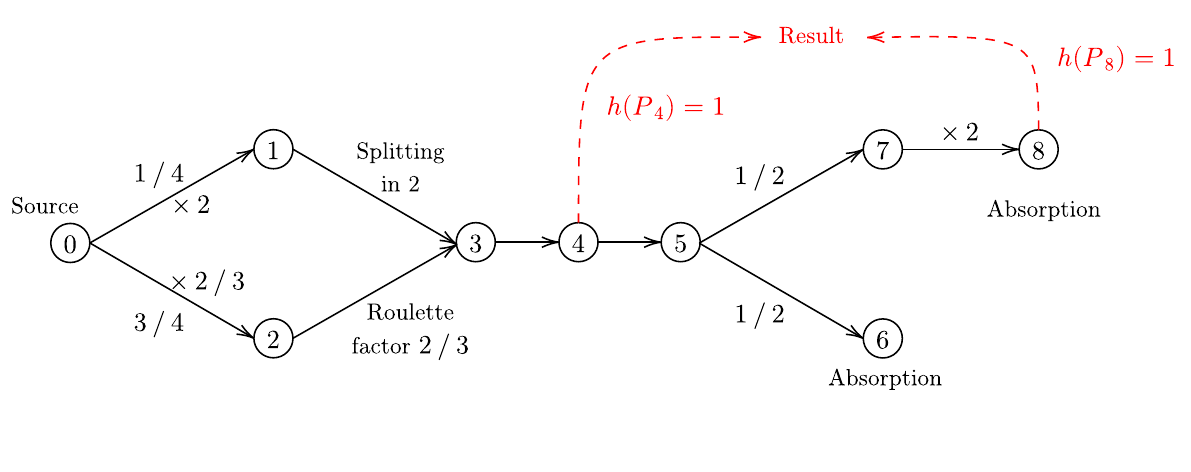}
    \caption{Example of a discrete Monte Carlo game composed of 9 states discrete states $P_i$. A single source particle is born in state $P_0$ (denoted simply 0 for brevity) with a unit weight. All particles then transition from state to state following the arrows and the adjacent probabilities. Along their paths, operations such as weight multiplication, roulette, splitting, or scoring can happen. A full description of the game is as follows: The particle is born in state 0 with unit weight. Then all particles in state 0 can either go to state 1 with probability $1/4$ and see their weight multiplied by 2 ($\times 2$ symbol) or go to state 2 with probability $2/3$ and see their weights multiplied by $2/3$. Particles in state 1 undergo splitting into 2 offsprings of half weight and go to state 3. Particles in state 2 undergo roulette and survive with probability $2/3$ seeing their weight multiplied by $3/2$, otherwise they die. Particles in state 3 go to state 4, score 1 (multiplied by their weights), then directly go to state 5. From there they then either go to state 7 with 50\% chance or die in state 6. Particles in state 7 see their weights multiplied by 2 and go to state 8 where they finally score $1\times$Weight. We will consider the case of one source particle, but this game could be played with an arbitrary number $N$ of independent source particles.}
    \label{fig:example_discrete}
\end{figure}

The first example involves a discrete Monte Carlo game that is simple enough to be solved analytically. The game is defined on a discrete ensemble of states $P_i$ in which the particles can be; transitions between the states with associated probabilities are assigned. In this example, a single source particle starts with unit weight at point $P_\text{source}=P_0$. Operations on the weights of the particles, such as a weight multiplication, roulette or splitting are included. The particles can score when present on some states according to an estimator function $h$, leading to a result $R$ whose variance we want to study. The game is illustrated in detail in \cref{fig:example_discrete}. This somewhat artificial problem, which bears no resemblance to particle-transport applications, is chosen so to better understand the variance-decomposition formula in a setting where reference solutions exist for comparison.

In order to investigate this problem, the first step is to analytically determine the solution of its direct and adjoint formulations. Solving the direct problem will yield the matter density, $\rho$ and the average particle density $n$. Note that as the problem at hand is discrete, the densities naturally use the discrete counting measure and can be interpreted as quantities at a given point (e.g. $n(P)$ represents the average quantity of particles at point $P$). The adjoint problem will yield the importance $M_1$. The obtained quantities are summarized in \cref{table:solution_dir_adj_ex_discrete}.

\begin{table}[htbp]
\centering

\begin{tabular}{||c||c c c c c c c c c||} 
 \hline
 States & $P_0$ & $P_1$ & $P_2$ & $P_3$ & $P_4$ & $P_5$ & $P_6$ & $P_7$ & $P_8$ \\ [0.6ex] 
 \hline
 \hline
 $\rho(P)$ & 1 & 1/2 & 1/2 & 1 & 1 & 1 & 1/2 & 1/2 & 1 \\ 
 \hline
 $M_1(P)$    & 2 & 2   & 2   & 2 & 2 & 1 & 0   & 2   & 1  \\ 
 \hline
 $n(P)$      & 1 & 1/4 & 3/4 & 1 & 1 & 1 & 1/2 & 1/2 & 1/2 \\ 
 \hline
 
\end{tabular}
\caption{Matter density $\rho$, importance $M_1$ and average number of particles $n$, in each discrete state $P_i$ of the game described in \cref{fig:example_discrete}.}
\label{table:solution_dir_adj_ex_discrete}
\end{table}

We will now derive the variance of the game described in \cref{fig:example_discrete}, for one source particle, using several methods: first we will derive it analytically by evaluating the different possibilities, then we will derive it using the variance-decomposition formula in \cref{eq:decomp_var}, and finally we will derive it using the moment equations. This will allow us to compare the nature of the variance-decomposition formula to that of the moment equations, and analyze how they relate to each other.

\subsubsection{Analytical derivation}

Let us derive analytically the probabilities of presence of the particles, which allows establishing the law of the result $R$. Let us denote $N_i$ the random variable describing the number of particles that pass at point $P_i$. The crucial state in \cref{fig:example_discrete} is 3. If the particle goes to state 1 at the beginning, it has been split in two particles, and $N_3 = 2$. Otherwise, the particle goes to state 2 and undergoes roulette, leading to either $N_3=0$ or $N_3 =1$. The probabilities for $N_3$ are then:
\begin{equation}
    \begin{aligned}
        \mathbb{P}(N_3=2)&=1/4\\
        \mathbb{P}(N_3=1)&=3/4 \times 2/3 = 1/2\\
        \mathbb{P}(N_3=0)&=3/4 \times 1/3 = 1/4.
    \end{aligned}
\end{equation}
The weight of particles in state 3 are always 1. We can then run the same analysis on state 7 conditionally on $N_3$: this allows us to derive directly the result $R$. For example, the probability of having $N_7=2$ if $N_3=2$ is $1/4$: both particles choose state $7$ over state $6$. The corresponding score is 2 from state 4, and $2\times 2$ for state $8$ (as the weights are multiplied by two between states 7 and 8), which yields $R=6$. This analysis is conducted for all possibilities: 
\begin{equation}
    \begin{aligned}
        \mathbb{P}(N_7=2 \cap N_3=2 )&=1/4 \times 1/4 =1/16 \quad  &(R=6)\\
        \mathbb{P}(N_7=1 \cap N_3=2 )&=1/2 \times 1/4 =1/8  &(R=4)\\
        \mathbb{P}(N_7=0 \cap N_3=2 )&=1/4 \times 1/4 =1/16 &(R=2)\\
        \mathbb{P}(N_7=1 \cap N_3=1 )&=1/2 \times 1/2 =1/4 &(R=3)\\
        \mathbb{P}(N_7=0 \cap N_3=1 )&=1/2 \times 1/2 =1/4  &(R=1)\\
        \mathbb{P}(N_7=0 \cap N_3=0 )&=1 \times 1/4   =1/4 &(R=0).
    \end{aligned}
\end{equation}
The second moment of $R$ is then
\begin{equation}
    \esp{R^2} = \frac{6^2}{16}  + \frac{4^2}{8}+ \frac{2^2}{16} + \frac{3^2}{4}  + \frac{1}{4} = 7.
    \label{eq:example_analytical_derivation_second_moment}
\end{equation}
The expected result being $\esp{R}=2$, we get a relative variance of 
\begin{equation}
    \relvar{R}=3/4.
    \label{eq:example_relvar}
\end{equation}

\subsubsection{Variance-decomposition formula}

The variance-decomposition formula decomposes the variance of the result over the different sampling occuring in the generalized phase space. As the studied Monte-Carlo game is discrete, the generalized phase space $\Pi$ is not continuous, and the use of the discrete counting measure turns the integral in \cref{eq:decomp_var} into a sum. 

The only points in $\Pi$ where an actual sampling occurs are in state 0 to choose the outgoing state, in state 2 for the roulette, and in state 5. Therefore, the relative variance of $R$ will be expressed as a sum over these states.

The sampling in state 0 experiences all the contribution, i.e. $c_r(P_0)=1$, the sampling intensity is $I_s(P_0)=1$ (all particles and all the contribution), and the variance of the weights is $\relvarw{w(P_0)}=0$; the variance of the weights in the whole simulation is also zero. The intrinsic variance is that of a binary process:
\begin{equation}
    \relvar{X_{P_0}}= (2 - 2/3)^2\frac{1}{4}\frac{1}{3}=1/3.
\end{equation}
The roulette sampling of state 2 experiences half of the total contribution, i.e. $c_r(P_2)=1/2$, and a sampling intensity $I_s(P_2)= 3/2$ ($3/4$ of the particles for $1/2$ of the contribution); its intrinsic variance is also that of a binary process:
\begin{equation}
    \relvar{X_{P_2}}= (3/2)^2\frac{2}{3}\frac{1}{3}=1/2.
\end{equation}
Finally, the sampling of state $5$ experiences half of the total contribution, i.e. $c_r(P_5)=1/2$ (half the total contribution has already been collected in state $4$), and a sampling intensity $I_s(P_5)=2$ ($n(P_5)=1$, while $c_r(P_5)=1/2$). The intrinsic variance is again that of a binary process:
\begin{equation}
    \relvar{X_{P_5}}= (2)^2\frac{1}{2}\frac{1}{2}=1.
\end{equation}
The final variance-decomposition formula yields then:
\begin{equation}
    \relvar{R} = \frac{c_r(P_0)}{I_s(P_0)}\relvar{X_{P_0}} + \frac{c_r(P_2)}{I_s(P_2)}\relvar{X_{P_2}} + \frac{c_r(P_5)}{I_s(P_5)}\relvar{X_{P_5}} = \frac{1}{3}+ \frac{1}{2}\frac{1}{\frac{3}{2}}\frac{1}{2}+\frac{1}{2}\frac{1}{2}1 = \frac{3}{4}.
\end{equation}
The formula allows recovering the exact result (see \cref{eq:example_relvar}), and at the same time decomposing the contributions over the distinct sources of variance in phase space.

\subsubsection{The moment equations}

It is useful to evaluate the expectation and the variance of the game using the moment equation formalism proposed by \citet{lux_monte_2018}. Generally speaking, the moment equations are constructed by setting a hierarchy of backward (adjoint) equations for the moment of order $k$ of the result. Each equation depends on the equations of the moments of orders up to $k-1$, which are supposed to be already solved: the hierarchy is thus finite at any given order.

The second moment function, denoted $M_2(P)$, yields in particular the expectation of $R^2$ for a particle of unit weight starting at $P$. The equation for $M_2(P)$ is expressed as a (backward) balance relating the second moment at a point $P$ to the second moment (and first moment) at other points $P'$ from which the point can move to $P$. Solving this equation yields the second moment $M_2(P)$ for all the phase space points $P$: we then compute $\esp{R^2}$ by evaluating $M_2(P)$ at the position of the source.

We remind the reader of the notation $R_i$ for a random variable corresponding to the result of a game in which a single unit weight particle is created at point $P_i$. Let us start with the second moment in state 8: we have $M_2(P_8) = 1$, since a single particle of unit weight placed at state $8$ would score $R_8^2=1$. To obtain the second moment in state 7, we need to take into account the weight multiplication. In full generality, as the second moment of a particle might depend on its weight in a non-trivial way (for example if population control depends on the entering weight of the particle), one might need to use an extended version of the second moment including the start weight $M_2(P,w)$. However, if no weight-dependent operations occur in the game (this is the case in our example), we simply have:
\begin{equation}
    M_2(P,w) = w^2M_2(P,1) = w^2M_2(P),
\end{equation}
and we can use the standard $M_2(P)$ function\footnote{The theory of the generalized moment equation has been derived in \citet{booth_analysis_1979}.}. In our case, the moment equation relating states 7 and 8 is
\begin{equation}
    M_2(P_7)= 2^2 M_2(P_8) = 4.
\end{equation}
Going back to state 5, a particle starting in $5$ will either die in state $6$, or go to state 7 with equal probability. The second moment equation is then a weighted average of the second moment of the different outcomes:
\begin{equation}
    M_2(P_5) = \frac{1}{2} M_2(P_6) + \frac{1}{2} M_2(P_7) = 2.
\end{equation}
To obtain the second moment at state 4, we decompose $R_4$:
\begin{align}
    M_2(P_4) &= \esp{R_4^2} \nonumber\\
    &= \esp{(h(P_4)+R_5)^2} \nonumber\\
    &= h(P_4)^2+ 2 h(P_4)\esp{R_5} + \esp{R_5^2} \nonumber\\
    &= h(P_4)^2 +2h(P_4) M_1(P_5) + M_2(P_5)\\
    &= 1 + 2 +2 = 5. \nonumber
\end{align}
State 3 is the same as state 4. State 2 is related to state 3 through roulette: it is a bifurcation between a path leading to death and a path leading to survival with probability 2/3 (with a multiplication factor 3/2). The second moment is then:
\begin{equation}
    M_2(P_2) = \frac{2}{3}\times \left( \frac{3}{2}\right)^2 M_2(3) = \frac{15}{2}.
\end{equation}
The state 1 is related to state 3 through splitting into two particles (that will then behave independently) of half-weight. Denoting $R_{3,1}$ and $R_{3,2}$ two independent random variables obeying the same law as $R_3$ (the two random variables represent the score of each offspring), we obtain
\begin{align}
    M_2(P_1) &= \esp{R_1^2} \nonumber\\
    &= \esp{ \left(R_{3,1}/2 + R_{3,2}/2 \right)^2}\nonumber\\
    &= \frac{1}{4}\left( \esp{R_{3,1}^2} + \esp{R_{3,2}^2} + 2 \esp{R_{3,1}}\esp{R_{3,2}}  \right)\nonumber\\
    &= \frac{1}{4} \left(2M_2(P_3) + 2 \left(M_1(P_3)\right)^2 \right)\\
    &= \frac{9}{2}\nonumber.
\end{align}
In a general case, if we split a particle in $m$ offsprings of weights $1/m$ in between $P$ and $P'$, the second moment equation is
\begin{equation}
    M_2(P) = \frac{1}{m}\left[ M_2(P') + (m-1)\left(M_1(P')\right)^2\right].
\end{equation}
Finally, the moment equation in state 0 is a weighted average:
\begin{equation}
    M_2(P_0) = \frac{1}{4}2^2M_2(P_1) + \frac{3}{4} \left( \frac{3}{2} \right)^2 M_2(P_2) = 7.
\end{equation}
As expected, since $\esp{R^2} = M_2(P_0)$, we recover the same result as in \cref{eq:example_analytical_derivation_second_moment}.

Comparing the moment equations to the variance-decomposition formula, we observe some peculiar differences. The equations for the second moments obey a system that must be solved in order to actually extract the $M_2(P)$. In our example, the particles cannot go back to a state they have visited before, and the corresponding moment equations yield a triangular system that can be simply solved iteratively for each state $P$. In the general case, the moment equations are fully coupled, and a linear solver would be in principle required to determine $M_2(P)$. This means in particular that the impact of a given variance-reduction technique might be difficult to understand, as the resolution of the system adds an intermediary step relating the chosen technique and its impact on $M_2(P)$ and hence on $\esp{R^2}$. On the contrary, the variance-decomposition formula is expressed as a simple integral, and the effects of a given variance-reduction technique can usually be interpreted more straightforwardly. 

Notwithstanding the previous remark, the moment equations convey more information than the decomposition formula, in that they determine the second moment for a particle starting anywhere in phase space; a similar analysis is precluded for the variance-decomposition formula (which yields by construction a global analysis), unless all the terms are recomputed.

\subsection{Examples involving continuous-energy transport in infinite media}

We focus now on realistic particle-transport problems. We start by considering continuous-energy transport in infinite media. In this configuration the samplings are limited to the source and collision events (flights are not explicitly represented). We choose the samplings $X_E$ to represent collisions starting at energy $E$; in addition, $X_\text{source}$ represents the source sampling. Therefore, the generalized phase space $\Pi$ is the energy range from $E_\text{min}$ to $E_\text{max}$, plus a single point $P_\text{source}$ that represents the state of a particle before the sampling of the source. The variance-decomposition formula in \cref{eq:decomp_var} takes the form:
\begin{equation}
    \relvar{R} = \frac{1}{N}\int_{E_\text{min}}^{E_\text{max}} c_r(E) \frac{1+\relvarw{w(E)}}{I_s(E)}\relvar{X_E}dE\;+ \frac{1}{N}\relvar{X_\text{source}},
    \label{eq:decomp_var_energy_0D}
\end{equation}
where the relative contribution of the source is $c_r(P_\text{soure})=1$, its sampling intensity is $I_s(P_\text{source})=1$ and its weight variance $\relvarw{w(P_\text{source})}=0$. This formula ignores the variance induced by any population-control technique: this variance will be taken into account in \cref{sec:example_U8H1} in \cref{eq:decomp_var_pop_control_measurable}.

In order to numerically evaluate the relative contribution density $c_r(E)$, we run a direct and adjoint Monte Carlo simulation. For the latter, the adjoint source is the detector of the direct simulation. We tally the collision density $\psi$ and the collision importance $\psi^\dagger$, and record the sought response $\esp{R}$ through either the result of the direct or the adjoint simulation. Technically, we only produce one sample of $R$, and use it as an estimator of the sought response $\esp{R}$. Similarly, the tallied $\psi$ and $\psi^\dagger$ are taken as estimators of the true collision density and collision importance. We can then compute $c_r(E)$ using the formula
\begin{equation}
    c_r(E) = \frac{\psi(E)\psi^\dagger(E)}{\esp{R}}.
\end{equation}
In order to evaluate the variance of the weights $\relvarw{w(E)}$, we can use the special variance estimators introduced in \citet{rovel_general_2025}, which measure the variance of all the statistical weights of the particles that have a collision in a small energy interval. In all the subsequent examples, this contribution to variance will be minimal, and will not play a significant role in the final variance. To evaluate the sampling intensity $I_s(E)$, we will use special estimators that count the number of particles (without taking their weights into account) to obtain an estimation of the average particle density $n(E)$, and then use the definition of $I_s$ given in \cref{eq:sampling_intensity} to obtain
\begin{equation}
    I_s(E) = \frac{n(E)}{Nc_r(E)}.
\end{equation}
The final and most difficult term to evaluate is the intrinsic variance of a collision. For a particular $E$ value, we send a `test particle' of weight $w=1$ into the sampling $X_E$, i.e., we sample one collision. Then, for all the $K$ offsprings (if any), we collect at the output the weight $w'_k$ and the importance of each particle $\chi^\dagger(E'_k)$. We can then compute a realization of the random variable $M_1(X_E)$ as (see definition in \cref{eq:def_importance_xp}):
\begin{equation}
     M_1(X_E):= h(P)+\sum_{k=1}^K m_k M_1(P'_k)=h(E)+\sum_{k=1}^K w'_k \chi^\dagger(E'_k).
    \label{eq:importance_of_sampling_energy_0D}
\end{equation}
By repeatedly sampling this collision $X_E$ with several test particles, we can get independent samples of $M_1(X_E)$ and evaluate the intrinsic variance by computing the relative variance of the samples of $M_1(X_E)$( See \cref{eq:intrinsic_variance}).

We can then probe the whole phase space by repeating this procedure on a representative mesh over the energy domain. The variance of the source will not be evaluated here and in all subsequent examples, since it will be shown to be negligible for our configurations.

\subsubsection{Purely elastic scattering}
\label{sec:example_pure_el_scat}

We consider first a medium allowing for a single isotropic elastic scattering reaction, with $A=6$. The viable energy domain is taken from $E_\text{min}= 0.1\text{ MeV}$ to $E_\text{max} = 20\text{ MeV}$. The source is placed uniformly in $[16.5, 18.1]\text{ MeV}$, and the source density is normalized to one source particle. The score collects collisions uniformly within $[0.11,0.12]\text{ MeV}$, and the detector response function $\eta_\psi$ is normalized so that its integrated response function amounts to 1~MeV (the adjoint source has a unit norm).

The direct and adjoint simulations are run using a Monte Carlo mini-app developed for the purpose of investigating adjoint Monte Carlo calculations; a thorough description of this tool is provided in \citet{rovel_general_2025}. The direct and adjoint calculations are run with $10^6$ particles, and the variance of the direct calculation is evaluated using batches composed of a single source particle to achieve maximum accuracy\footnote{The precision of the variance estimator increases with the number of realizations, i.e. with the number of batches. The maximum number of batches is obtained with single-particle batches.}. All functions of $E$ are measured by 2000 logarithmically-spaced bins spanning the whole energy range. The intrinsic variance estimation is done by sending 2000 test particles in each of the 2000 points over which it is evaluated.

\begin{figure}[htbp]
    \centering
    \begin{subfigure}[t]{0.32\textwidth}
        \centering
        \caption{Collision density $\psi(E)$}
        \includegraphics[width=1.1\textwidth]{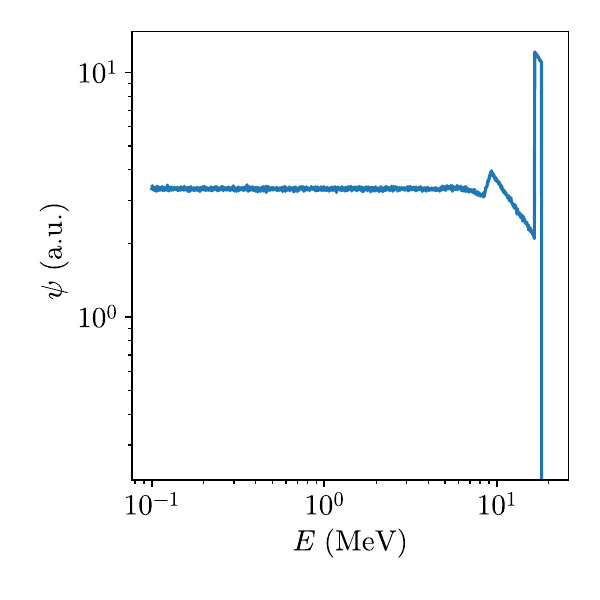}
        \label{fig:example_pure_el_scattering_drr}
    \end{subfigure}%
    ~
    \begin{subfigure}[t]{0.32\textwidth}
        \centering
        \caption{Emitted importance $\chi^\dagger(E)$}
        \includegraphics[width=1.1\textwidth]{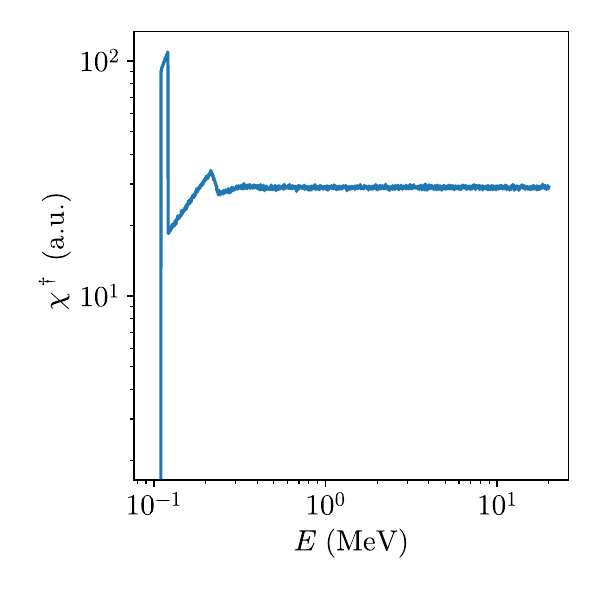}
        \label{fig:example_pure_el_scattering_aer}
    \end{subfigure}%
    ~
    \begin{subfigure}[t]{0.32\textwidth}
        \centering
        \caption{Contribution density $c_r(E)$}
        \includegraphics[width=1.1\textwidth]{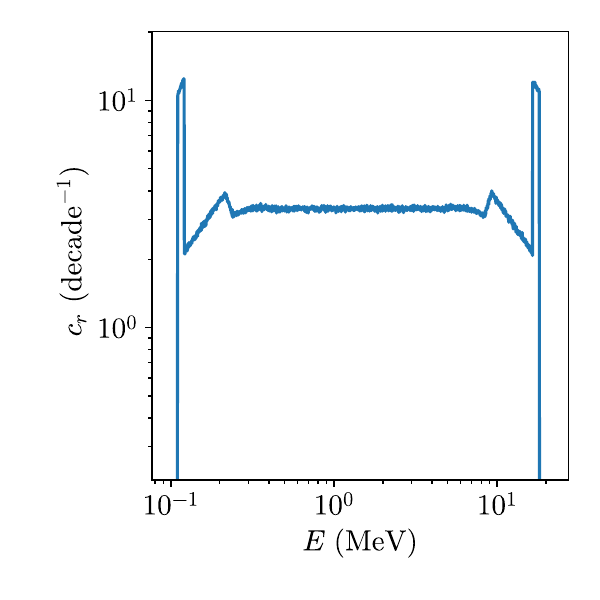}
        \label{fig:example_pure_el_scattering_rc}
    \end{subfigure}
    
    \begin{subfigure}[t]{0.32\textwidth}
        \centering
        \caption{Sampling intensity $I_s(E)$}
        \includegraphics[width=1.1\textwidth]{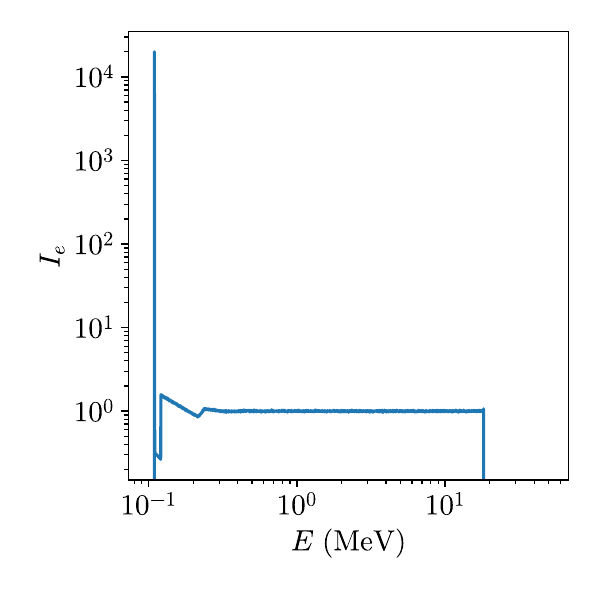}
        \label{fig:example_pure_el_scattering_si}
    \end{subfigure}%
    ~
    \begin{subfigure}[t]{0.32\textwidth}
        \centering
        \caption{Intrinsic variance $\relvar{X_E}$}
        \includegraphics[width=1.1\textwidth]{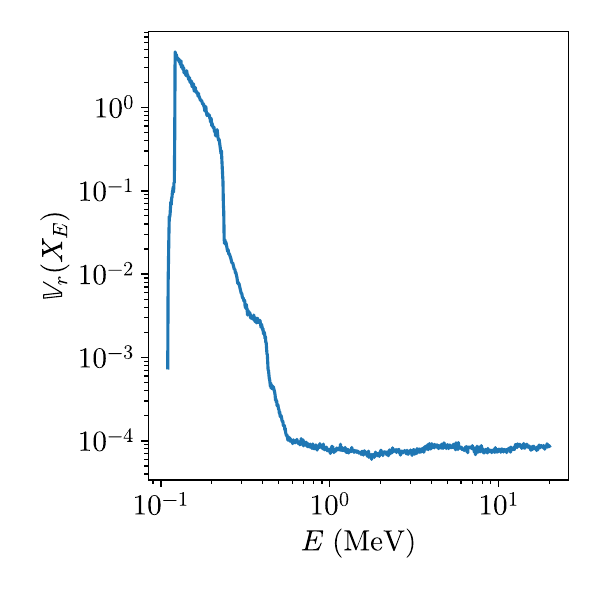}
        \label{fig:example_pure_el_scattering_iv}
    \end{subfigure}%
    ~
    \begin{subfigure}[t]{0.32\textwidth}
        \centering
        \caption{Distribution of the variance}
        \includegraphics[width=1.1\textwidth]{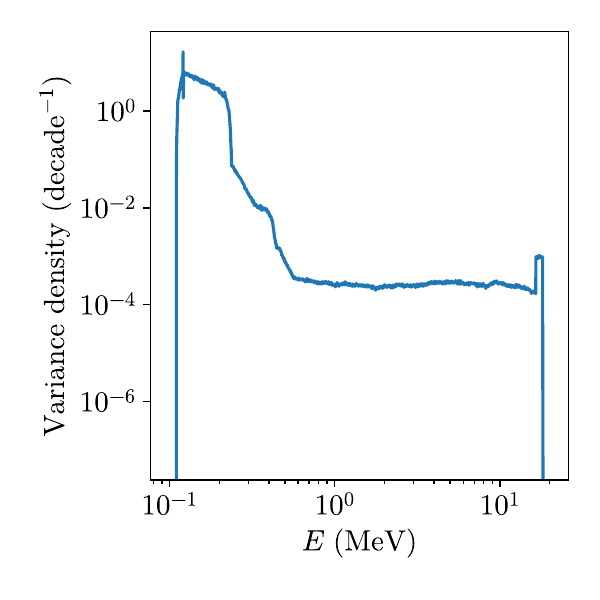}
        \label{fig:example_pure_el_scattering_vd}
    \end{subfigure}
    \caption{Data extracted from the direct and adjoint calculations of a purely elastic infinite medium with a source and detectors placed around $20$~MeV and $0.1$~MeV. See text for detailed description of the problem studied. Densities (including collision density, contribution density and distribution of the variance) are represented per decade of energies, hence the absence of the characteristic $1/E$ shape of the flux in a slowing down problem. The detailed analysis of the results is provided in the main text.}
    \label{fig:example_pure_el_scattering}
\end{figure}

The different simulation results and data extracted to evaluate the variance-decomposition formula are presented in \cref{fig:example_pure_el_scattering}. The shapes of the collision density $\psi$ (see \cref{fig:example_pure_el_scattering_drr}) and emission importance $\chi^\dagger$ (see \cref{fig:example_pure_el_scattering_aer}) display the classical Placzek and inverse Placzek transients. The contribution density $c_r(E)$ (see \cref{fig:example_pure_el_scattering_rc}), which is the product of $\psi$ and $\psi^\dagger = \chi^\dagger$, shows that the contribution emerges from the source where it is very concentrated; it then `spreads' in the energy range, while `traveling' to lower energies; finally, it concentrates again into the `detector'. The sampling intensity (see \cref{fig:example_pure_el_scattering_si}) is constant close to the source, and varies close to the detector: far from the detector, the importance is flat, and the paths followed by the particles are also the natural paths of the contribution: the relative contribution is proportional only to the direct collision density. However, close to the detector, the relative contribution converges towards the detector, whereas the particles keep uniformly decreasing in energy, following the analog slowing-down process. This leads to some points, especially in the detector region, being under-sampled: this yields a state with high relative contribution, but average number of particles actually sampling it. At energies lower than that of the detector, the contribution is null, since no particles can be up-scattered. However, the analog particles still sample these energies, which leads to a diverging sampling intensity. The shape of the intrinsic variance (see \cref{fig:example_pure_el_scattering_iv}) is extremely interesting: it is rather small at energies far away from the detector, but rises at energies just above it, and in particular at energies from which the detector is reachable in a single collision. This can be understood using the importance in \cref{fig:example_pure_el_scattering_aer}, which is flat far away from the detector and strongly varies near and within the detector region. Thus, all the exit points for collisions far away from the detector have roughly the same importance, and the variable $M_1(X_E)$ is approximately constant, with very small relative variance. On the other hand, when the detector zone is within the range of accessible energies for a given collision, the importance of the exit states has wild variations: e.g.~sampling an energy within the detector yields a large importance, while sampling an energy beyond the detector yields a zero importance; the intrinsic variance is thus very large.

Finally, we examine the overall distribution of the variance (see \cref{fig:example_pure_el_scattering_vd}), which originates from the intrinsic variance and all its pre-factor terms (corresponding to the integrand in \cref{eq:decomp_var_energy_0D}): the analysis of this quantity shows where the total variance is generated in phase space. The largest contribution to the variance stems from the collisions near the detector, since this is a zone with a high intrinsic variance $\relvar{X_E}$, and high relative contribution $c_r(E)$.

When the detector region is `small' in the phase space, typically the intrinsic variance (and thus the final variance) is mainly due to the last few samplings before reaching the detector. In this case, the last flights and/or collisions actually determine whether the particle will land in the detector and score, or outside and possibly never score. We have systematically observed this feature in several other test-cases (not reported in this paper). This finding sheds new light on \cref{eq:weight_target_precise}, and suggests that increasing the sampling in regions where the intrinsic variance is higher should be beneficial for the FoM. Even though the intrinsic variance at every sampling is not precisely known, in order to improve the FoM we can slightly increase the sampling in a region near the detector (e.g. by lowering the target weight of a weight window)\footnote{Using a target weight inversely proportional to the importance, as suggested in \cref{eq:weight_target_simple}, already increases the sampling near the detector; however, the suggested improvement would increase the sampling even more, to take into account the increased intrinsic variance as suggested by \cref{eq:weight_target_precise}}.

The variance induced by the source (that is not evaluated here) is very small, since all the source outgoing states, i.e.~all the possible source outgoing energies, have the same importance: the source lies in the region where the importance is very uniform. This is exacerbated by the fact that the range of energies accessible via the source ($[16.5, 18.1]\text{ MeV}$) is quite narrow in comparison to the range of energies accessible via one collision.

Integrating the variance density to obtain the final variance, using \cref{eq:decomp_var_energy_0D} and neglecting the variance of the sampling of the source, yields a  unitary (for a single source particle) relative variance of $2.742$. Running the actual direct Monte Carlo simulation yields a unitary relative variance of $2.767 \pm 0.007$ (here and in the following, statistical uncertainties will be expressed as 1-$\sigma$ standard deviation). The variance is measured with maximal precision using batches consisting of a single source particle; the variance of the variance (VoV) is quantified by computing the variance over 100 independent estimates of the variance. The predicted variance is close to the 3-$\sigma$ uncertainty interval of the exact variance, despite all the possible sources of (small) discrepancies, encompassing the following: we have neglected the variance of the source; the quantities estimated during the direct and adjoint Monte Carlo simulations might be poorly converged, especially concerning the intrinsic variance; furthermore, inaccuracies might be introduced by the energy mesh not being fine enough. Converging the intrinsic variance estimate can be quite difficult, since the variance depends on the second moment, and its convergence is slower than usual average quantities.

\subsubsection{Adjoint simulations with realistic nuclear data}
\label{sec:example_U8H1}

We illustrate then the behavior of the variance-decomposition formula for an example concerning adjoint particle transport in a mixture composed of equal concentrations of \textsuperscript{238}U and \textsuperscript{1}H. Nuclear data are taken from the ENDF-B/VIII.0 library, at $k_BT=0.1 \text{ eV}$. This example problem is inspired by similar benchmark configurations previously considered in \citet{rovel_preparing_2026}.

The (direct) source is placed uniformly in [16.5,18.1]~MeV, and the detector (i.e., the adjoint source) is placed uniformly in [0.11,0.12]~eV. Their respective norms are fixed to unity, like in the previous example. The viable energy domain is bounded by $E_\text{min}=10^{-4}\text{ eV}$ and $E_\text{max}=20\text{ MeV}$. Similarly to the first example, the energy range is divided into 2000 logarithmically-spaced energy bins. 

The direct and adjoint Monte Carlo simulations are run with $10^6$ particles organized into batches of a single particle, for maximum-accuracy variance measurements. In order to evaluate the intrinsic variance, the importance $M_1$ of the outgoing states is the collision density $\psi$: the problem adjoint to an adjoint problem is the direct problem.

As detailed in \citet{rovel_general_2025}, simulating adjoint transport via Monte Carlo involves a substantial amount of weight corrections. To prevent the divergence of the weights, a weight window is used. The variance induced by this population-control process can be numerically measured, and we will later show that this variance contribution is small compared to the others. 

To obtain the full variance induced by population control on the whole energy range, we can simply integrate \cref{eq:var_pop_control} over the whole energy range. Obtaining terms that are numerically measurable from \cref{eq:var_pop_control} can be difficult. However, by using the definition of the sampling intensity, of the contribution and of the average weight (see \cref{eq:sampling_intensity}, \cref{eq:def_rel_contrib} and \cref{eq:def_avg_weight_from_rho}, respectively), the equation for the variance over the whole energy range can be expressed as
\begin{align}
    &\var{\text{Pop. Control}}\nonumber\\
    &=\frac{1}{N} \int_{E_\text{min}}^{E_\text{max}} \frac{c_r(E)}{I_s(E)}\int_{w=0}^\infty \frac{n(E,w)}{n(E)}\frac{(w_t(w))^2}{\espw{w(E)}^2} (f-\lfloor f \rfloor)(\lfloor f+1\rfloor-f) dwdE\nonumber\\
    &=\frac{1}{N} \int_{E_\text{min}}^{E_\text{max}} \frac{(M_1(E))^2}{N\left( \esp{R}\right)^2}\int_{w=0}^\infty n(E,w)(w_t(w))^2 (f-\lfloor f \rfloor)(\lfloor f+1\rfloor-f) dwdE.
    \label{eq:decomp_var_pop_control_measurable}
\end{align}
The term $M_1$ in \cref{eq:decomp_var_pop_control_measurable} corresponds to $\psi$, for the meaning of the terms related to population control, refer to \cref{sec:pop_control}. The expected value of the result is estimated from either direct or adjoint simulations. Finally, the integral over the weight can be estimated using a custom tally that records the quantity $(w_t(w))^2 (f-\lfloor f \rfloor)(\lfloor f+1\rfloor-f)$  at every population-control event.
\begin{figure}[htbp]
    \centering
    \begin{subfigure}[t]{0.32\textwidth}
        \centering
        \caption{Collision density $\psi(E)$}
        \includegraphics[width=1.1\textwidth]{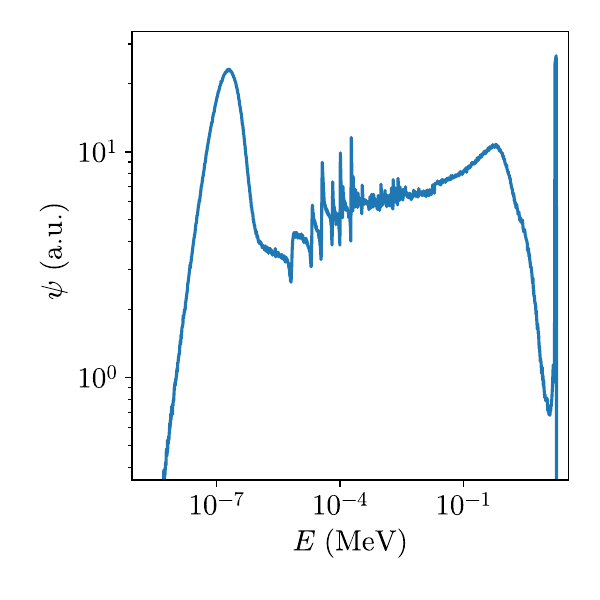}
        \label{fig:example_U8H1_drr}
    \end{subfigure}%
    ~
    \begin{subfigure}[t]{0.32\textwidth}
        \centering
        \caption{Emission importance $\chi^\dagger(E)$}
        \includegraphics[width=1.1\textwidth]{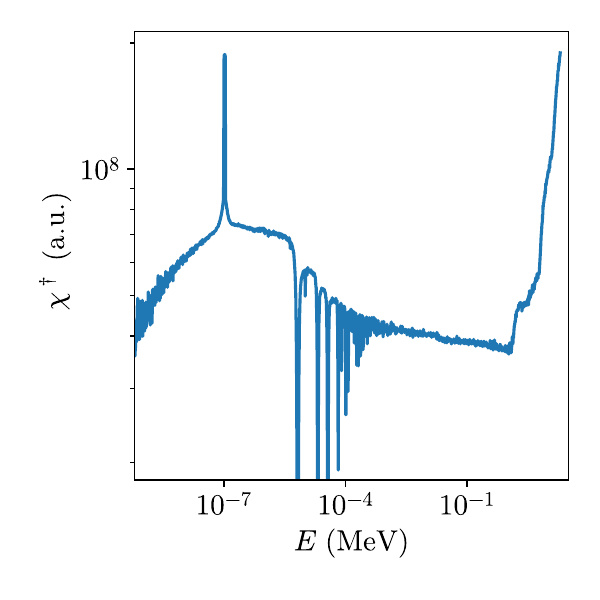}
        \label{fig:example_U8H1_aer}
    \end{subfigure}%
    ~
    \begin{subfigure}[t]{0.32\textwidth}
        \centering
        \caption{Contribution density $c_r(E)$}
        \includegraphics[width=1.1\textwidth]{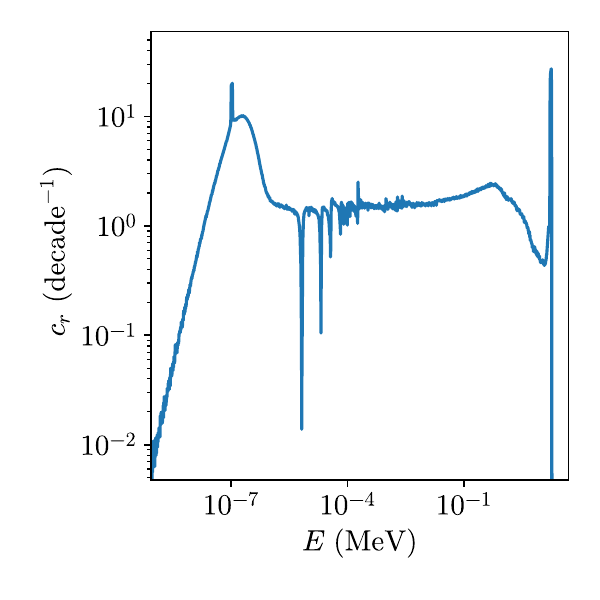}
        \label{fig:example_U8H1_rc}
    \end{subfigure}
    
    \begin{subfigure}[t]{0.32\textwidth}
        \centering
        \caption{Cumulated collision variance}
        \includegraphics[width=1.1\textwidth]{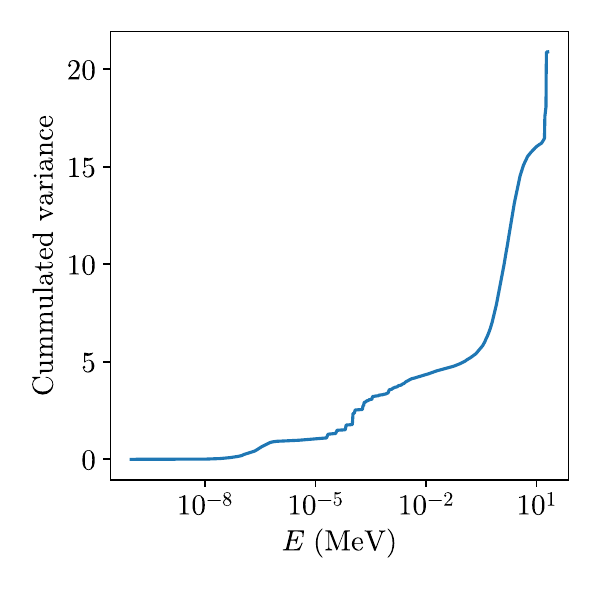}
        \label{fig:example_U8H1_cummulated_var}
    \end{subfigure}%
    ~
    \begin{subfigure}[t]{0.32\textwidth}
        \centering
        \caption{Pop. control variance}
        \includegraphics[width=1.1\textwidth]{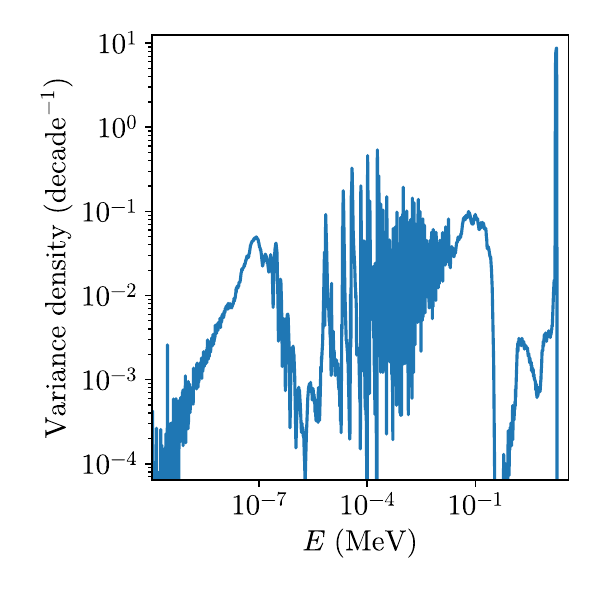}
        \label{fig:example_U8H1_weight_control_variance}
    \end{subfigure}%
    ~
    \begin{subfigure}[t]{0.32\textwidth}
        \centering
        \caption{Pop. cont. cumulated variance}
        \includegraphics[width=1.1\textwidth]{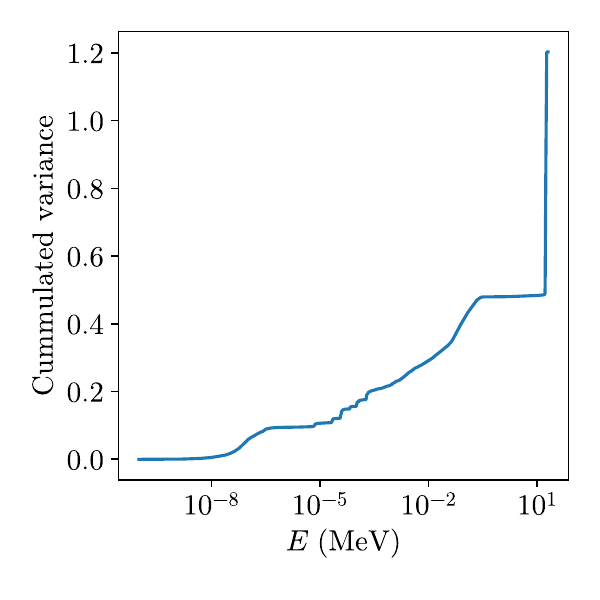}
        \label{fig:example_U8H1_weight_control_cum_var}
    \end{subfigure}
    \caption{Data extracted from the direct and adjoint calculations of a \textsuperscript{238}U and \textsuperscript{1}H equimolar infinite medium, with source and detector placed around $20$~MeV and $0.1$~eV. A detailed description is provided in the main text. The collision density, contribution density and the distribution of the variance are represented per decade of energies, which explains the absence of the characteristic $1/E$ shape of the flux in a slowing-down problem. The variances evaluated using the variance-decomposition formula in \cref{eq:decomp_var} are evaluated for the adjoint Monte Carlo game using the detector as adjoint source, and the source as adjoint detector. Cumulated variance graphs (in \cref{fig:example_U8H1_cummulated_var,fig:example_U8H1_weight_control_cum_var}) express variance densities integrated over $E$, to better pinpoint the variance-inducing regions in phase space. The analysis of the results in provided in the main text.}
    \label{fig:example_U8H1}
\end{figure}

The collected data and the computed variance decomposition are presented in \cref{fig:example_U8H1}. The collision density $\psi(E)$ (see \cref{fig:example_U8H1_drr}) displays a Maxwellian shape at thermal energies, a slowing-down pattern with absorption resonances, and finally a fission spectrum around 1~MeV. The source at 18~MeV is located at the far right of the plot. The emission importance $\chi^\dagger(E)$ (see \cref{fig:example_U8H1_aer}) displays the presence of the detector at 0.1~eV, the resonances and a sharp increase once the fission energy threshold of \textsuperscript{238}U is reached, at around 1.5~MeV. The relative contribution density $c_r(E)$ (see \cref{fig:example_U8H1_rc}) shows the combined effects of $\psi(E)$ and $\chi^\dagger(E)$: the relative contribution is concentrated in the source, spreads across energies between source and detector, and concentrates again in the detector. The sampling intensity $I_s$, the intrinsic variance $\relvar{X_E}$ and the weight variance $\relvarw{w(E)}$ are not shown. The final variance distribution is shown in \cref{fig:example_U8H1_cummulated_var}, integrated over (energy) phase space. Recall that the variance evaluated here corresponds to an adjoint simulation, where the particles start from the detector and score upon reaching the source. The cumulative distribution singles out the effects of very high but very localized variance contributors (such as resonances). The analysis of \cref{fig:example_U8H1_cummulated_var} shows that a small portion of the variance is added when the adjoint particles exit the thermal region, then when they cross the resonance region\footnote{A thorough analysis of the variance-inducing nature of the resonances for adjoint simulations is provided in \citet{rovel_general_2025}.}, but most of the variance is added at high energies, once adjoint fission reactions can lead directly to the adjoint detector located around 18~MeV. The total variance added by the collisions amounts to a relative unitary variance of  about 20.9. \Cref{fig:example_U8H1_weight_control_variance,fig:example_U8H1_weight_control_cum_var} show that the weight window used to control the dispersion of the statistical weights induced by the adjoint scheme adds variance of its own. However, it amounts to a total of a relative unitary variance of about 1.2, i.e. small compared to the variance induced by collisions.

The variance spike added by the population control in the adjoint detector region represents roughly half of the total variance added by population control. This spike is not due to discretization errors, nor is the weight window particularly active in this small region. \Cref{eq:var_pop_control} shows the variance induced by a population-control procedure and allows concluding that the term $c_r(P)/I_s(P)$ is very large in the detector region, due to large relative contribution and low sampling intensity. Thus, the rather mild effect of the weight window in the detector region is amplified by this term, leading to a large impact on the variance. This suggests that suppressing population control in the detector might prove slightly beneficial when the detector region is narrow (which leads to a large relative contribution and a dip in the sampling intensity). 

The total variance obtained through the variance-decomposition formula includes the component stemming from collisions (about 20.9) and the component stemming from population control (about 1.2), and amounts to about 22.1. Again, the variance stemming from source sampling is neglected, as it is very narrow ($[0.11,0.12]$~eV) and all outgoing energies have roughly the same importance $\psi$. This value is to be compared with the actual unitary relative variance measured during the adjoint simulation, which reads $19.6\pm0.12$. This time, the value from the variance-decomposition formula lies clearly outside the statistical uncertainty range. This is likely due to a poorly converged phase-space meshing: the 2000 logarithmically-spaced energy bins are most likely insufficient to capture the complex structure of the resonances. However, the variance-decomposition formula allows obtaining a correct order of magnitude for the variance, which is a rather satisfactory achievement.

\subsection{A two-dimensional single-speed transport problem}
\label{sec:example_streaming}
\begin{figure}[ht]
    \centering
    \includegraphics[width=.7\textwidth]{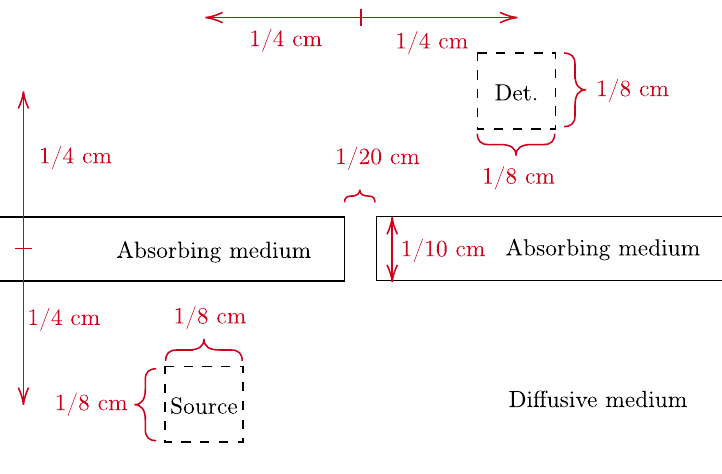}
    \caption{Geometry of the two-dimensional single-speed transport problem. A source and a detector are located in a diffusive medium, separated by an absorbing wall where a small hole allows particles to stream from the source region to the detector region. No boundary conditions are imposed. The source and the detector are taken uniform and of unit norm. The source is composed of collided particles, and we use a collision estimator. The diffusive medium has a scattering cross section $\Sigma_s = 10\text{ cm}^{-1}$ and an absorption cross section $\Sigma_a = 0.1\text{ cm}^{-1}$. Scattering is isotropic. The absorption medium has a scattering cross section $\Sigma_s = 1\text{ cm}^{-1}$ and an absorption cross section $\Sigma_a = 100\text{ cm}^{-1}$. The wall is 0.1 cm thick and has a width of $10\Sigma_t^{-1}$, ensuring that the probability of crossing it is negligible. The typical diffusion length in the diffusive medium is $(\Sigma_a \Sigma_s)^{-1/2} = 1$~cm, which is close to the distance between the source and the detector.}
    \label{fig:streaming_example_geom}
\end{figure}

We focus next on a space-dependent transport problem that allows examining the contribution of flights, and in particular particle streaming effects, to the variance. The details of the geometrical model are illustrated in \cref{fig:streaming_example_geom}. A source and a detector region are placed in a scattering medium, and are separated by a strongly absorbing wall. A narrow channel through the wall allows particles to cross between the two adjacent regions. We assume single-speed transport and isotropic scattering, and we restrict the setting to a two-dimensional configuration.

We take the elementary sampling process $X_P$ to be a full cycle of a collision followed by a flight. To ensure that the particle history can be decomposed into events of the kind $X_P$, the source for this process is taken to be the first-collision source: the source particles are generated directly in the first-collision events.

In full generality, evaluating the variance-decomposition formula for this problem would require a mesh over the entire phase space, including two dimensions for the position and one dimension for the direction. However, the isotropic nature of the scattering term allows us to decompose the phase space only along the two spatial dimensions. This is possible because the collision/flight cycle is similar for all entering directions (since the collision is isotropic), and the importance of its outcomes (the collisional importance $\psi^\dagger$) is also isotropic. The variance-decomposition formula is then expressed as in the infinite-medium continuous-energy example, the integral over energy being replaced by a two-dimensional integral over space. Thus, \crefrange{eq:decomp_var_energy_0D}{eq:importance_of_sampling_energy_0D} can be straightforwardly adapted by replacing the energy variable $E$ by a position variable $(x,y)$.

Although the problem presented in \cref{fig:streaming_example_geom} has an infinite spatial extent, we will score the different quantities on a mesh of $100\times100$ square bins that span a total of 1~cm $\times$ 1~cm. The adjoint and direct simulations are run with a total of $10^7$ particles, organized into batches of a single source particle to achieve the best possible convergence of the variance estimator. The VoV of the direct simulation is estimated by comparing 1000 independent variance estimations. Crossing the absorbing boundary is an unlikely event: to improve the convergence of the direct and adjoint simulations, a narrow weight window (with an opening of $1.1$) is therefore used, with a target weight proportional to the inverse of the importance of the problem\footnote{We use the inverse of the direct collision density $\psi$ as a weight target for the adjoint calculation, and the inverse of the collisional importance $\psi^\dagger$ as a weight target for the direct calculation. The weight-window process is applied before collision events for the direct simulations, and after the adjoint collision events for the adjoint simulations.}. This means that we approximately enforce a sampling intensity of $1$ everywhere in the problem. To evaluate the intrinsic variance of the collision/flight events, at each of the $100\times100$ spatial bins $2\times10^6$ test particles of unit weight are sampled isotropically in a state where they directly enter collision. Their importance is then measured after a collision/flight cycle, using the collision importance $\psi^\dagger$. If the test particle exits the mesh of $1\text{ cm}\times1\text{ cm}$ where the importance $\psi^\dagger$ has been measured, we take the importance of the closest point on the mesh.

\begin{figure}[htbp]
    \centering
    \begin{subfigure}[t]{0.46\textwidth}
        \centering
        \caption{Collision density $\psi(x,y)$ (a.u.)}
        \includegraphics[width=1.1\textwidth]{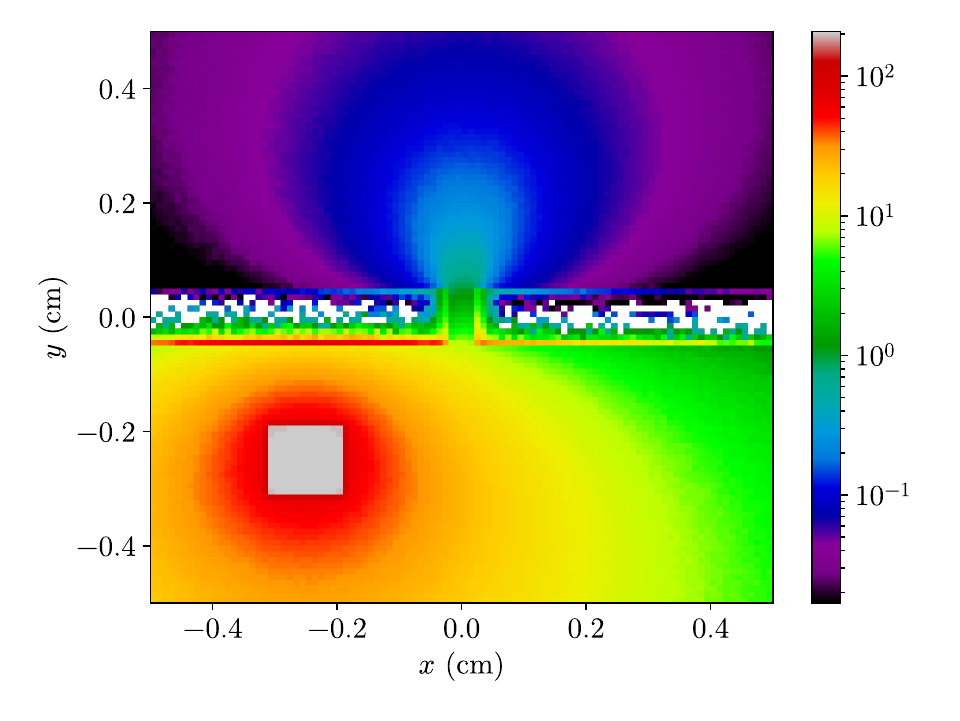}
        \label{fig:example_streaming_drr}
    \end{subfigure}%
    ~
    \begin{subfigure}[t]{0.46\textwidth}
        \centering
        \caption{Collision importance $\psi^\dagger(x,y)$ (a.u.)}
        \includegraphics[width=1.1\textwidth]{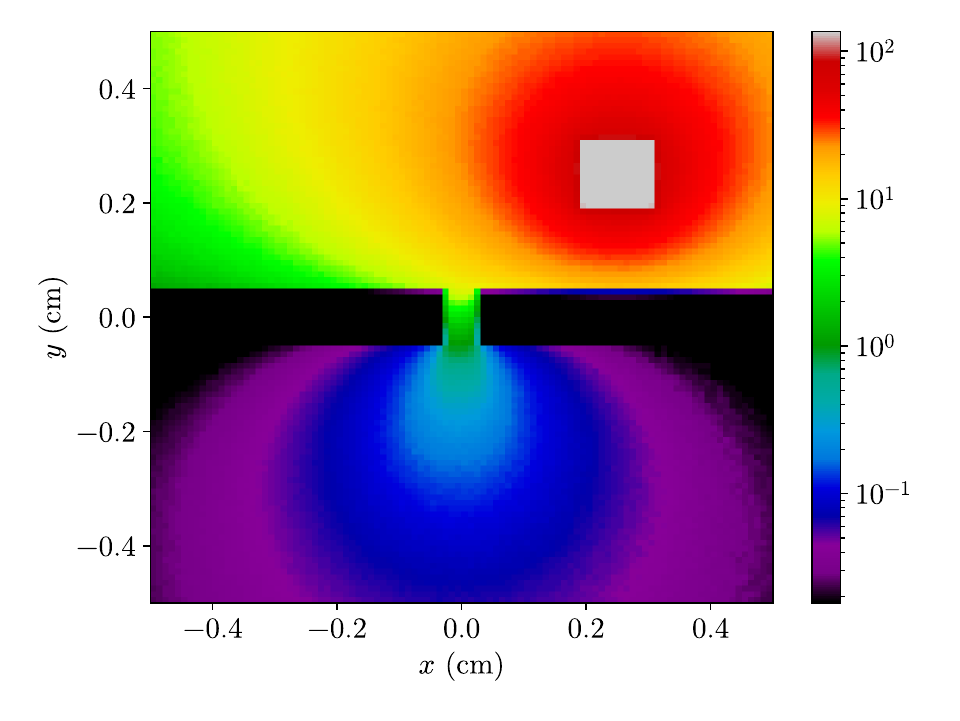}
        \label{fig:example_streaming_aer}
    \end{subfigure}
    
    \begin{subfigure}[t]{0.46\textwidth}
        \centering
        \caption{Contribution density $c_r(x,y)\;(\text{cm}^{-2})$}
        \includegraphics[width=1.1\textwidth]{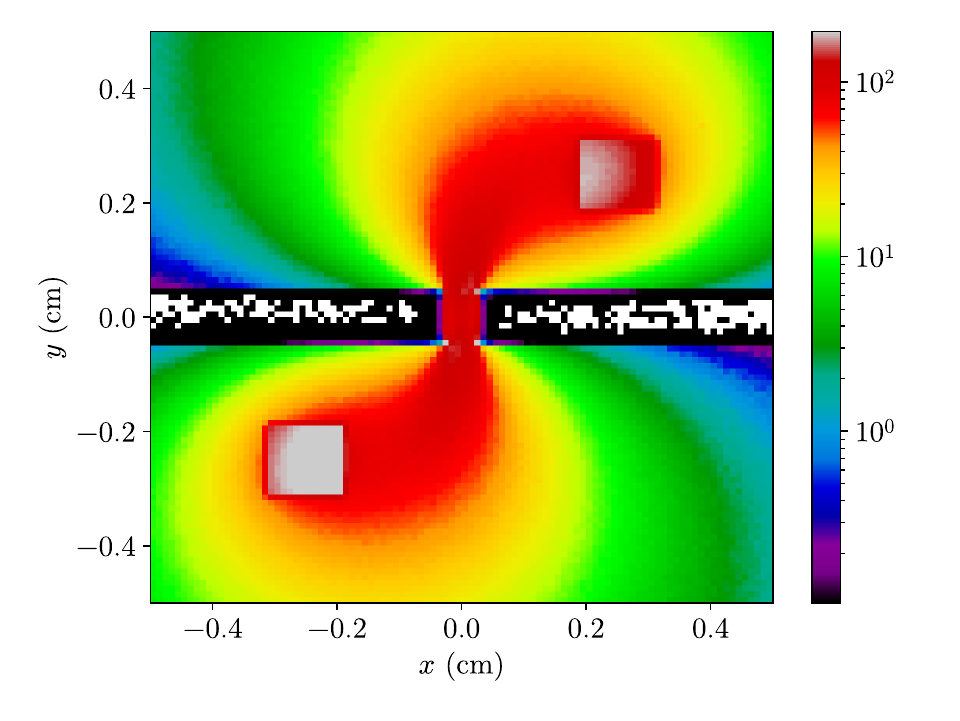}
        \label{fig:example_streaming_rc}
    \end{subfigure}%
    ~
    \begin{subfigure}[t]{0.46\textwidth}
        \centering
        \caption{Inverse sampling intensity $I_s(x,y)^{-1}$}
        \includegraphics[width=1.1\textwidth]{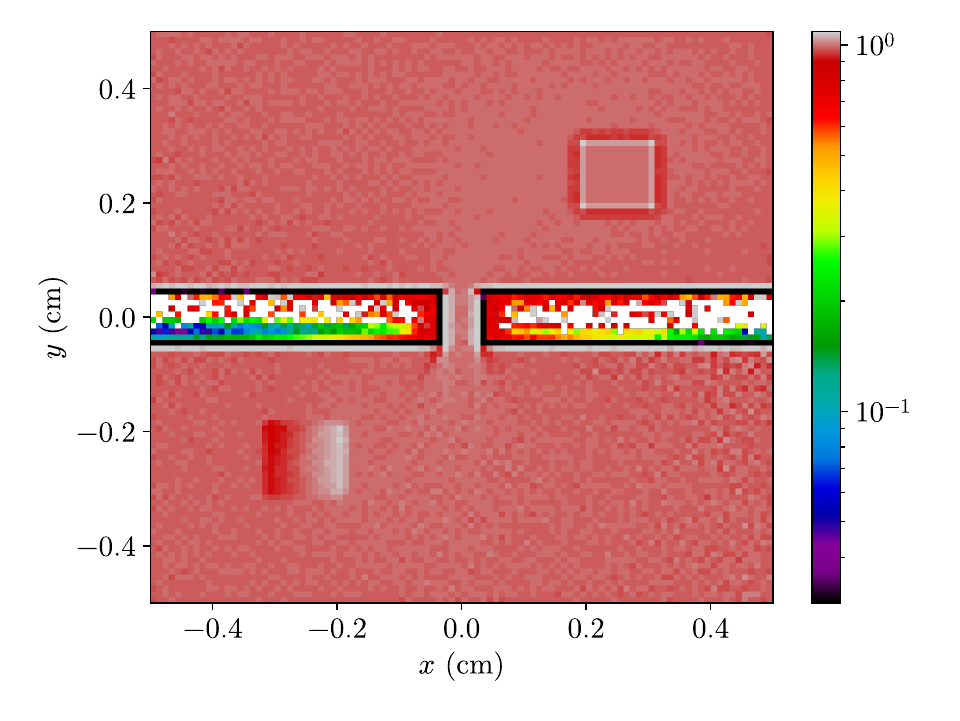}
        \label{fig:example_streaming_isi}
    \end{subfigure}
    
    \begin{subfigure}[t]{0.46\textwidth}
        \centering
        \caption{Intrinsic variance $\relvar{X_P}$}
        \includegraphics[width=1.1\textwidth]{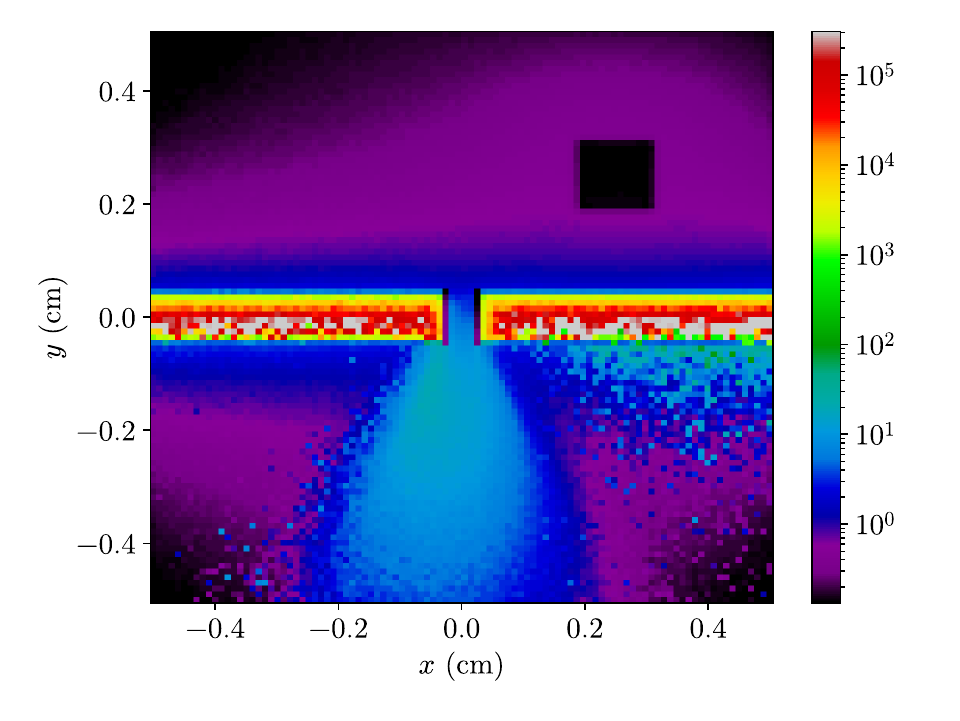}
        \label{fig:example_streaming_iv}
    \end{subfigure}%
    ~
    \begin{subfigure}[t]{0.46\textwidth}
        \centering
        \caption{Distribution of the variance $(\text{cm}^{-2})$}
        \includegraphics[width=1.1\textwidth]{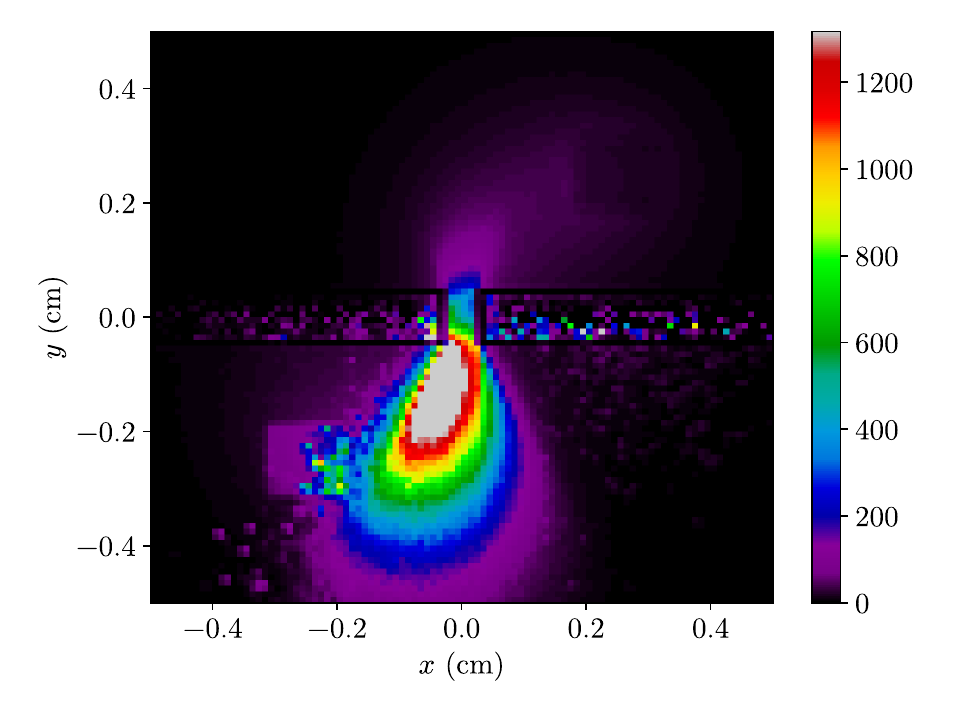}
        \label{fig:example_streaming_vd}
    \end{subfigure}
    \caption{Data extracted from the direct and adjoint simulations of the streaming problem presented in \cref{fig:streaming_example_geom}. The problem description and the analysis of the results are provided in the main text.}
    \label{fig:example_streaming}
\end{figure}

The results of the simulations are presented in \cref{fig:example_streaming}. The direct collision density $\psi$ (see \cref{fig:example_streaming_drr}) shows that the particles spread through the diffusive medium and can cross to the other side of the absorbing boundary using the streaming hole. The collision importance $\psi^\dagger$ (see \cref{fig:example_streaming_aer}), shows a similar pattern, with the importance being large in the diffusive medium around the detector, and reaching the other side via the streaming hole. The contribution density $c_r$ (see \cref{fig:example_streaming_rc}) shows that the contribution that flows from the source to the detector passes through the streaming hole. The amount of contribution crossing the absorption wall directly (i.e. without streaming through the hole) is negligible. \Cref{fig:example_streaming_drr,fig:example_streaming_aer,fig:example_streaming_rc} display an overall symmetry between the adjoint and direct simulations. This symmetry is only approximate, as the source is composed of collided particles, while the estimator collects collided particles: the adjoint source creates adjoint particles that will first undergo adjoint flights. \Cref{fig:example_streaming_isi} shows $I_s^{-1}$, the inverse of the sampling intensity: the strict imposed weight window has the intended effect of ensuring a roughly constant sampling intensity of 1 everywhere in the problem. The only exception is deep within the absorbing wall, where the target weight was not properly converged. This has no practical effect, since contributions do not pass through the absorbing wall. The relative weight variance (not shown here) is negligible, as the weight window enforces a strict control of the weights. 

\Cref{fig:example_streaming_iv} displays the intrinsic variance $\relvar{X_P}$: the intrinsic variance tends to increase the closer we get to the absorbing wall. This can be understood as follows. In the diffusive medium, the main factor influencing the future importance of the particle is whether the particle dies (which is more likely in the absorbing wall) or survives. The intrinsic variance is maximal deep within the absorbing wall, because a particle entering collision here has a tiny chance of not being absorbed and flying out of the wall; this makes a non-zero importance a very rare event, leading to a large relative variance. Also, particles that are facing the streaming hole (and have not yet crossed it) have a large intrinsic variance: such particles have a low probability of being scattered in the right direction and crossing the absorbing boundary to reach the other side (where the importance is orders of magnitude higher). Furthermore, converging the estimator of the variance is even more difficult when rare events are included, which explains the noise occurring in \cref{fig:example_streaming_iv}. Finally, taking into account all the pre-factors in the variance-decomposition formula in \cref{eq:decomp_var}, we get the variance distribution over phase space shown in  \cref{fig:example_streaming_vd}: the variance is mainly generated by the collision and flight that occur in front of the streaming hole. The very large intrinsic variance occurring in and around the wall is not relevant for the variance density, as its relative contribution is very weak, contrary to the variance due to crossing the hole, which is amplified by all the contribution flowing through it.

Integrating the variance distribution over phase space, we obtain the final relative variance of the result $R$. Again, we neglect the variance of the source sampling and the variance induced by the population control procedure\footnote{The variance of population control is often small in comparison with the variance of flights and collisions, as shown in the previous example \cref{sec:example_U8H1}. The variance of the source is only due to the birth positions of the particles, as their first action is to undergo an isotropic collision. As seen in \cref{fig:example_streaming_aer}, the importance over the source zone is only varying mildly (roughly $\pm20\%$) in comparison with the variations seen across the wall (at least a factor 100).}. The variance-decomposition formula yields a unitary relative variance of about 82, which is compatible with the variance $83\pm 0.7$ estimated by Monte Carlo simulations. 

The variance is mainly generated by collisions/flights that start in a relatively limited region in phase space. This suggests a rather simple way of improving the FoM. By manually lowering the weight target in this region, we locally increase the sampling intensity, and mechanically decrease the impact of this region on the final variance. 

\begin{figure}[htbp]
    \centering
    \begin{subfigure}[t]{0.46\textwidth}
        \centering
        \caption{Inverse sampling intensity $I_s(x,y)^{-1}$}
        \includegraphics[width=1.1\textwidth]{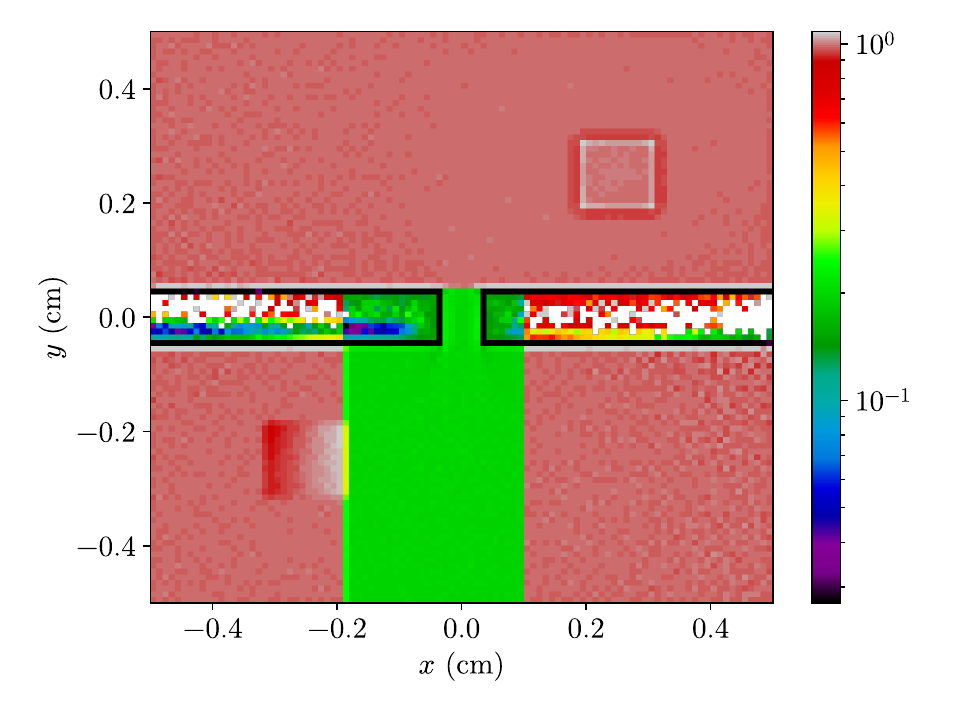}
        \label{fig:example_streaming2_isi}
    \end{subfigure}%
    ~
    \begin{subfigure}[t]{0.46\textwidth}
        \centering
        \caption{Distribution of the variance $(\text{cm}^{-2})$}
        \includegraphics[width=1.1\textwidth]{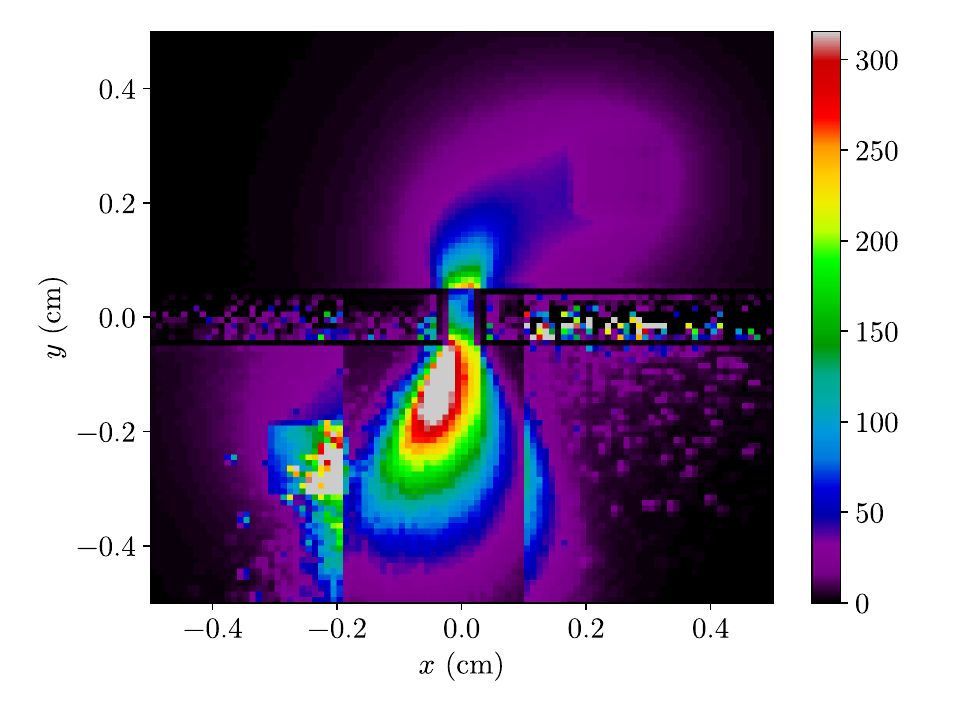}
        \label{fig:example_streaming2_vd}
    \end{subfigure}
    \caption{Data extracted from the direct and adjoint calculations of the streaming problem presented in \cref{fig:streaming_example_geom}. Adapted weight window in used to increase the sampling intensity in the zone in front of the streaming hole by a factor $5$. Only data differing from \cref{fig:example_streaming} is shown. See text for detailed simulation explanation and result analysis.}
    \label{fig:example_streaming2}
\end{figure}

For the example considered here, we decreased the weight window of the direct calculation by a factor of $5$ in the region $x\in[-0.1875,0.1]\text{ cm}$ and $y\in[-0.5,0.05]\text{ cm}$. The effects of this modified simulation are presented in \cref{fig:example_streaming2}. The inverse $I_s^{-1}$ of the sampling intensity is displayed in \cref{fig:example_streaming2_isi}: we now have a distinct region where the sampling intensity is 5 times larger. The effect on the variance density is shown in \cref{fig:example_streaming2_vd}: the variance generated in the region is reduced by a factor 5. Once integrated, the variance-decomposition formula yields a unitary relative variance of about 31, which is compatible with the actual variance $32\pm 0.5$ estimated by Monte Carlo simulations. Adapting the weight window allowed reducing the variance by a factor of 2.7. The computing time for the direct simulation also grew from 62 seconds to 83 seconds, since the number of collisions and flights increased in the region. The FoM was then improved by a factor of about 2.0. The final variance, shown in \cref{fig:example_streaming2_vd},  remains concentrated in phase space: we might further adapt the weight window towards the optimal solution given by \cref{eq:weight_target_precise}, which would further increase the FoM.

\section{Conclusions}

In this work we presented a new formula to analyze the variance produced by direct or adjoint  Monte Carlo games and decompose its overall value into several contributions. This approach complements the use of the well-known moment equations, by providing a formula that is easier to wield and used to assess variance-reduction techniques. The obtained expression for the final relative variance of a given result takes the form of an integral of the individual intrinsic variances generated by the sampling events met by the simulated particle histories along the game, weighted by their relative importance, and thus allows conceptually attributing the final variance of the sought result to the single components of the game. A full demonstration of the formula has been provided in \cref{sec:proof}.

After presenting the general structure of the variance-decomposition formula and detailing the meaning of its terms, we have illustrated its application to the interpretation of commonly used variance-reduction techniques (encompassing implicit capture, branchless simulation, weight windows, importance sampling or path stretching). We have focused in particular on the assessment of population-control methods, and more generally the issue of how the computation time should be best allocated on different regions of phase space in order to achieve optimal performance. In this respect, FoM-optimal games have been introduced, and a formula capable of probing their performance in terms of computing time and obtained variance (the inverse of the Figure of Merit) has been derived, using a suitable ersatz for the amount of time spent in a game.

Based on these considerations, the variance-decomposition formula has been then used to obtain a different derivation of zero-variance for direct transport problems and obtain the rules of zero-variance for adjoint transport problems.

Finally, a series of benchmark configurations have been analyzed to verify the validity of the formula and illustrate the application of said formula to practical Monte Carlo games. The considered examples range from simple discrete-state problems where analytical solutions are available to complex continuous-energy adjoint transport problem using real nuclear data. Several insights regarding how variance is generated in Monte Carlo games have been obtained, and the variance-decomposition formula has been compared to the moment equations.

The results provided in this work  are a stepping stone towards the improvement of existing variance-reduction techniques or even the formulation of new ones. Building upon our findings, future research work might concern the derivation of a general numerical approach capable of assessing the terms appearing in the variance-decomposition formula. This would allow for both an estimation of the variance before running a Monte Carlo game, and the derivation of tailored variance-reduction techniques (in a similar way as in \cref{sec:example_streaming}) yielding increased efficiency. Another useful generalization of the variance-decomposition formula would concern the possibility of taking into account games allowing for weights having arbitrary signs, such as those emerging in neutron noise problems \citep[see e.g. ][]{fauvel_convergence_2025}, which would lead to a novel approach to the investigation of their associated statistical dispersion.

\appendix
\crefalias{section}{appendix}

\renewcommand{\theequation}{\thesection.\arabic{equation}}

\setcounter{equation}{0}
\section{Using different estimators}
\label{sec:non_col_estimators}

In the problem setting (\cref{sec:context_and_def}) we restricted ourselves to estimator function $h$ that only depend on a single point $P$. At first glance, this seems incompatible with some estimators such as flight based estimators denoted $h_f(\mathcal{P},\mathcal{P'})$ that depend on both the starting point of a flight $\mathcal{P}$ and its end point $\mathcal{P'}$. For this reason we restricted ourselves to collision estimators in the main part of this work. In this appendix, we will focus on the effect of using different estimators when solving transport problems with Monte Carlo.

The problem of the dependence of flight based estimators on points $\mathcal{P}$ and $\mathcal{P'}$ can actually be circumvented easily. By considering that, in the generalized phase space $\Pi$, we incorporate `transition points' $P$ that represent the state of a particle in flight from $\mathcal{P}$ to $\mathcal{P'}$. For example the sampling of the flight then gets decomposed into two sub-steps: at the beggining of the flight, the particle is at a point (that we will denote $P_0$) that represents the particle being emitted at $\mathcal{P}$. Then, the flight is sampled (the collision point $\mathcal{P'}$ is chosen), and the particle goes to point $P_1$ that represents the flight from $\mathcal{P}$ to $\mathcal{P'}$. In that state we can use the estimator function $h(P_1)=h_f(\mathcal{P},\mathcal{P'})$. Then the particle is sent to $P_2$ that represents the particle being collided at $\mathcal{P'}$ (this last step is a deterministic sample).

Using the method described above, transport problems with generic estimators fit into the problem setting described in \cref{sec:context_and_def}, and the variance decomposition formula applies. Lifting the restriction to use collision estimators, however, has some impact on the zero-variance formulas derived in \cref{sec:zero_variance}. In the rest of this section we will analyze these impacts. 

\subsection{Boltzmann equation and importance}
\label{sec:boltz_eq_and_importance}

A difficulty that emerges when trying to apply the general framework described in \cref{sec:context_and_def} to transport problems that use generic estimators stems from the nature of the importance $\psi^\dagger$ and $\chi^\dagger$ as defined by the adjoint transport equation. This will lead to $M_1(P)$ not being equal to $\psi^\dagger$ when $P$ is a collision point, modifying the formulas seen in \cref{sec:zero_variance}.

In the classical framework of the transport equation, (see for example \citet{lux_monte_2018} for a complete introduction to this framework), we compute the collision density $\psi$ and the emission density $\chi$ from the source density $S$ using
\begin{equation}
    \begin{aligned}
        \chi(\mathcal{P}) &= \int C(\mathcal{P}'\to \mathcal{P})\psi(\mathcal{P}')d\mathcal{P}' + S(\mathcal{P})\\
        \psi(\mathcal{P}) &= \int T(\mathcal{P}'\to \mathcal{P})\chi(\mathcal{P}')d\mathcal{P}',
    \end{aligned}
    \label{eq:boltz_equation}
\end{equation}
with $T$ the flight kernel, turning the emission density $\chi$ into the corresponding collision density $\psi$ and $C$ the collision kernel turning the collision density $\psi$ into the re-emission density. From there, the sought response $\esp{R}$ can be obtained either via a collision based response function $\eta_\psi$ or via an emission based response function $\eta_\chi$ via:
\begin{equation}
    \esp{R}=\int \psi(\mathcal{P})\eta_\psi(\mathcal{P})d\mathcal{P} = \int \chi(\mathcal{P})\eta_\chi(\mathcal{P})d\mathcal{P},
\end{equation}
where one can compute $\eta_\chi$ from $\eta_\psi$ using 
\begin{equation}
    \eta_\chi(\mathcal{P})=\int T(\mathcal{P}\to \mathcal{P}')\eta_\psi(\mathcal{P}')d\mathcal{P}'.
    \label{eq:link_eta_psi_eta_chi}
\end{equation}
This leeway in the choice of detector response function has an impact on the definition of the adjoint boltzmann equation that defines the importance $\psi^\dagger$ and $\chi^\dagger$. Using a collision-based response function $\eta_\psi$ as adjoint source (which is most often the case) yields the following adjoint boltzmann equation
\begin{equation}
    \begin{aligned}
        \chi^\dagger(\mathcal{P}) &= \int T(\mathcal{P}\to \mathcal{P}')\psi^\dagger(\mathcal{P}')d\mathcal{P}' \\
        \psi^\dagger_{\eta_\psi}(\mathcal{P}) &= \int C(\mathcal{P}\to \mathcal{P}')\chi^\dagger(\mathcal{P}')d\mathcal{P}'+ \eta_\psi(\mathcal{P}),
    \end{aligned}
    \label{eq:adj_boltzmann_source_psi}
\end{equation}
while using an emission-based response function yields
\begin{equation}
    \begin{aligned}
        \chi^\dagger(\mathcal{P}) &= \int T(\mathcal{P}\to \mathcal{P}')\psi^\dagger(\mathcal{P}')d\mathcal{P}'+ \eta_\chi (\mathcal{P}) \\
        \psi^\dagger_{\eta_\chi}(\mathcal{P}) &= \int C(\mathcal{P}\to \mathcal{P}')\chi^\dagger(\mathcal{P}')d\mathcal{P}'.
    \end{aligned}
    \label{eq:adj_boltzmann_source_chi}
\end{equation}
Where we introduced an index $\eta_\chi$ or $\eta_\psi$ on the collision importance $\psi^\dagger$ to denote the adjoint source used. Note that in both cases $\chi^\dagger$ is identical, hence the absence of index on it. We can derive the following relation:
\begin{equation}
    \psi^\dagger_{\eta_\psi}=\psi^\dagger_{\eta_\chi}+\eta_\psi.
    \label{eq:link_importances}
\end{equation}
The `classical' definition of $\psi^\dagger$ (which we used in the main body of this work) is $\psi^\dagger_{\eta_\psi}$.

When trying to solve the Boltzmann equation via Monte Carlo, one uses estimator functions that can be very diverse as long as they respect the condition of being `partially unbiased' (for a thorough derivation of this condition see Chap. 5 Sec. V.A of \citet{lux_monte_2018}). We will state this condition for a `generic enough' estimator that scores $h_f(\mathcal{P},\mathcal{P'})$ during a flight from $\mathcal{P}$ to $\mathcal{P'}$ and scores $h_c(\mathcal{P'})$ upon collision at $\mathcal{P'}$. This is not fully general as one might score when specific reaction paths are taken (such as last event estimators) but is general enough to underline the problem at hand. The `partial unbiasedness' condition ensures that the importance $M_1(P)$ at the beginning of a flight corresponds to $\chi^\dagger$. The condition boils down to the fact that, on average over a full cycle flight+collision, the Monte Carlo game scores the same as a `natural game' that would use a collision estimator with estimator function $\eta_\psi$ :
\begin{equation}
    I:=\int T(\mathcal{P}\to \mathcal{P'})\left(h_f(\mathcal{P},\mathcal{P'}) + h_c(\mathcal{P'}) \right)d\mathcal{P'} = \int T(\mathcal{P}\to \mathcal{P'})\eta_\psi(\mathcal{P'}) d\mathcal{P'}.
    \label{eq:partially_unbiased_requirement}
\end{equation}
However, \cref{eq:partially_unbiased_requirement} cannot ensure on its own that $M_1(P)$ corresponds to $\psi^\dagger_{\eta_\psi}$ or $\psi^\dagger_{\eta_\chi}$ when $P$ is a collision point. $M_1(P')$ at a collision point $\mathcal{P'}$ can be computed as:
\begin{equation}
    M_1(P') = h_c(\mathcal{P'})+\int C(\mathcal{P'}\to \mathcal{P''})\chi^\dagger(\mathcal{P''})d\mathcal{P''}.
\end{equation}
Therefore, if one uses only a collision estimator ($h_f =0$), then \cref{eq:partially_unbiased_requirement} imposes that $h_c=\eta_\psi$ and $M_1(P)$ follows \cref{eq:adj_boltzmann_source_psi} leading to $M_1(P)=\psi^\dagger_{\eta_\psi}(\mathcal{P})$. Conversely, if one uses only a flight estimator ($h_c=0$), then $M_1(P)$ follows \cref{eq:adj_boltzmann_source_chi} and $M_1(P)=\psi^\dagger_{\eta_\chi}$.

A similar reasoning can be conducted on adjoint transport and adjoint Monte Carlo. One can derive two emission densities $\chi_{S_\chi}$ and $\chi_{S_\psi}$ depending if the adjoint detector (the source) $S$ is taken as an emission source $S_\chi$ or a collision source $S_\psi$. The partially unbiased condition only ensures that $M_1(P)=\psi(\mathcal{P})$ when the point $P$ represents a state before undergoing adjoint flight. If one uses a colision estimator, then one has $M_1(P)=\chi_{S_\chi}(\mathcal{P})$, whereas if one uses a flight estimator, then one has $M_1(P)=\chi_{S_\psi}(\mathcal{P})$; with the following relation:
\begin{equation}
    \chi_{S_\chi} = \chi_{S_\psi}+S_\chi.
\end{equation}

\subsection{Direct zero-variance formulas}
In this section we will re-obtain the zero-variance formulas from \cref{sec:zv_direct}. Let us begin with the treatment of the flight sample. Let us consider the case of a flight-based estimator $h_f(\mathcal{P},\mathcal{P'})$. As we introduced a special point $P_1$ that represents the flight from $\mathcal{P}$ to $\mathcal{P'}$ we need to compute its importance $M_1(P_1)$. As we use a flight-based estimator, the importance of a particle collided on $\mathcal{P'}$ is $\psi^\dagger_{\eta_\chi}=\psi^\dagger_{\eta_\psi}-\eta_\psi$. The importance of $P_1$ further adds the immediate payoff $h(P_1)$ yielding:
\begin{equation}
    M_1(P_1) = \psi^\dagger_{\eta_\psi}(\mathcal{P'})-\eta_\psi(\mathcal{P'})+h_f(\mathcal{P},\mathcal{P'}).
\end{equation}
This modifies the zero-variance formula (\cref{eq:zv_flight_ker}) into

\begin{equation}
    \hat{T}(\mathcal{P}\to \mathcal{P}') \propto T(\mathcal{P}\to \mathcal{P}') \left( \psi^\dagger_{\eta_\psi}(\mathcal{P'})-\eta_\psi(\mathcal{P'})+h_f(\mathcal{P},\mathcal{P'})\right),
\end{equation}

The sampling of of the collision event can be impacted if one uses an estimator function that depends on the reaction channel $\{i,j\}$ (nuclide $i$ reaction $j$). For example with an estimator $h_{i,j}(\mathcal{P'})$ that scores once the channel $\{i,j\}$ has been choosen (Note that this is different from the collision estimator $h_c$ introduced above that scored at the collision, before choosing the path). This kind of estimator is used in last event estimators that score only on capture events. Using it will modify the formula of the collided importance on nuclide $i$ reaction $j$ : $\psi^\dagger_{i,j}$. \cref{eq:importance_channel_ij} becomes:
\begin{equation}
    \psi^\dagger_{i,j} (\mathbf{r},E,\mathbf{\Omega}) =  h_{i,j}(\mathbf{r},E,\mathbf{\Omega})+\iint \nu_{i,j}(E)f_{i,j}(E\to E', \mathbf{\Omega}\cdot \mathbf{\Omega'})\chi^\dagger(\mathbf{r},E',\mathbf{\Omega'})d\mathbf{\Omega'}dE'.
\end{equation}
We can then choose the channel $\{i,j\}$ using the previously obtained \cref{eq:zv_choice_channel}. The rest of the sampling of the collision is unaffected.

\subsection{Adjoint zero-variance formulas}
In this section we will re-obtain the zero-variance formulas from \cref{sec:zv_adjoint}, in the case when we use a flight-based estimator $h_f(\mathcal{P'},\mathcal{P})$. As we introduced a special point $P_1$ that represents the adjoint flight from $\mathcal{P'}$ to $\mathcal{P}$ we need to compute its importance $M_1(P_1)$. As we use a flight-based estimator, the importance of an adjoint particle entering adjoint collision on $\mathcal{P}$ is $\chi_{S_\psi}=\chi_{S_\chi}-S_\chi$. The importance of $P_1$ further adds the immediate payoff $h(P_1)$ yielding:
\begin{equation}
    M_1(P_1) =\chi_{S_\chi}(\mathcal{P})-S_\chi(\mathcal{P})+h_f(\mathcal{P'},\mathcal{P}).
\end{equation}
This modifies the zero-variance formula (\cref{eq:zv_flight_ker}) into

\begin{equation}
    \hat{T}^\dagger(\mathcal{P}'\to \mathcal{P}) \propto T^\dagger(\mathcal{P}'\to \mathcal{P})  \left( \chi_{S_\chi}(\mathcal{P})-S_\chi(\mathcal{P})+h_f(\mathcal{P'},\mathcal{P}) \right).
\end{equation}

\setcounter{equation}{0}
\section{Proof of the variance-decomposition formula}
\label{sec:proof}

\subsection{The law of total variance}

Before addressing the proof of the variance decomposition formula given in \cref{eq:decomp_var}, we present the law of total variance, upon which the proof is based.

A fundamental result of probability theory, known as the `Law of total variance' (or the `variance-decomposition formula'), aims at expressing the variance of a random variable $A$ in terms of the variance of another random variable $B$; \citep[see e.g. ][Section 4.4]{casella_statistical_2024}. Formally, this law states that, if $A$ and $B$ are two random variables and $A$ has a finite variance, then the variance of $A$ can be decomposed as:
\begin{equation}
    \var{A} = \var{\esp{A|B}}+\esp{\var{A|B}}.
    \label{eq:law_total_var}
\end{equation}

The term $\esp{A|B}$ is the expectation of $A$ conditioned to a given outcome of $B$. Thus, the first term in \cref{eq:law_total_var} quantifies the variation of the (conditional) expected values of $A$ due to the outcomes of $B$, i.e.~the contribution to the variance of $A$ that can be explained by $B$. The term $\var{A|B}$ describes the variance of $A$ for a given outcome of $B$: this is the contribution to the variance of $A$ that cannot be explained by $B$. Since there exist many different possible outcomes of $B$, the total variance of $A$ not explained by $B$ is the expected value of the remaining variance for all outcomes of $B$, which yields the second term in \cref{eq:law_total_var}.

The proof of \cref{eq:law_total_var} is as follows:
\begin{align*}
    \var{A} & = \esp{A^2}- \esp{A}^2\\
     & = \esp{\esp{A^2|B}} - \big(\esp{\esp{A|B}}\big)^2 \\
     & = \esp{\var{A|B}+\esp{A|B}^2} - \big(\esp{\esp{A|B}}\big)^2\\
     & = \esp{\var{A|B}}+ \esp{\esp{A|B}^2} - \big(\esp{\esp{A|B}}\big)^2\\
     & = \esp{\var{A|B}} + \var{\esp{A|B}}.
\end{align*}

This formula can be generalized using $M$ `intermediary' random variables $B_i$, $i=1,\cdots,M$, to decompose the final result:
\begin{equation}
    \var{A} = \sum_{i=1}^M \esp{\var{\esp{A | B_1 , \cdots, B_i } | B_1, \cdots, B_{i-1} }} + \esp{\var{A|B_1,\cdots,B_M}},
    \label{eq:law_total_var_gen}
\end{equation}
for a given (deterministic) $M$.

Let us apply the law of total variance to the random variable $A$ that describes the result of a given Monte Carlo game. If $B_i$ are all the distinct sampling events that compose the (possibly branching) Markov chain associated to the game, then $A$ is completely defined by the $B_i$, and the conditional variable $A|B_1,\cdots,B_M$ is deterministic. The last term in \cref{eq:law_total_var_gen} is then zero, and the variance of $A$ is decomposed along the different sampling events $B_i$ composing its history. The $i$-th term is an average over all possible outcomes of $B_1,\cdots,B_{i-1}$ of the variance induced by $B_i$, i.e., the variance of the expected $A$ if events $B_i$ are sampled.

\subsection{Multi-particle Monte Carlo games}
\label{sec:multi_part_MC_proof}

We now set out to apply the generalized law of total variance to a full Monte Carlo game. We work with the setting defined in \cref{sec:context_and_def}. A first idea to decompose the variance of $R$ on the different samplings $X_P$ in phase space would be to apply the law of total variance (\cref{eq:law_total_var_gen}) using the samplings $X_P$ as `explanatory variables'. A problem with this approach is the dependence of the sampling $X_P$ at a certain point $P$ in the simulation on all the previous samplings. For example, a particle might be killed at different stages of the simulation. This means that a specific sampling $X_P$ might even not `happen' in some configurations.

To overcome this issue, we will decompose the variance over sampling variables $Y_i$ that are defined `sequentially'. In other words, the first sample occurring in the game is determined by $Y_1$, and so on. Thus, depending on the outcome of $Y_1,\cdots,Y_i$, the precise `meaning' of the sampling $Y_{i+1}$ might be totally different. For instance, in a given game $Y_{i+1}$ might represent the sampling of a flight, in another it might represent the sampling of a source to create a new particle, and in another game it might even have no impact at all because all the particles are already simulated, and the game is already finished. If we want $R$ to be entirely determined by the $Y_i$s, so that the last term in \cref{eq:law_total_var_gen} is zero, we need to have as many sampling events $Y_i$ as those possible in the Monte Carlo game, i.e. (in full generality) an arbitrarily large number. In a feasible Monte Carlo game, the probability that a history never dies is zero: the probability that the $(Y_i)_{i>M}$ have any influence over $R$ goes to zero as $M$ goes to infinity. The decomposition over $Y_i$ is useful, as it allows us to define a precise `state' $S_i$ of the simulation at step $i$, which corresponds to 
\begin{equation}
    S_i = (Y_1,\cdots,Y_i).
    \label{eq:def_state_i}
\end{equation}
By extension, the term $S_0$ represents the initial state of the simulation before any sampling has taken place. Using the generalized law of total variance with respect to $Y_i$ yields the following infinite sum:
\begin{equation}
    \var{R} = \sum_{i=1}^\infty \esp{\var{\esp{R | S_{i} } | S_{i-1} }}.
    \label{eq:total_var_chrono}
\end{equation}
The $i^\text{th}$ term in \cref{eq:total_var_chrono} represents the variance contributed by $Y_i$ (as once $S_{i-1}$ is fixed, $S_i$ is determined by $Y_i$ via \cref{eq:def_state_i}). The terms in the sum in \cref{eq:total_var_chrono} will not be practical to manipulate, since the meaning $Y_i$ will be obscure. Nonetheless, the decomposition formula being expressed as a sum, we will be able to regroup the terms based on their initial points $P$ to obtain an integral over phase space $\Pi$.

Let us now try to unveil the meaning of the terms occurring in \cref{eq:total_var_chrono}. The term $S_{i-1}$ imposes a current state of the Monte Carlo game. Once $S_{i-1}$ is fixed, we can interpret $Y_i$ as a specific sampling of a given particle that will be subsequently called the \emph{current particle}. We denote $P_{i-1}=P(S_{i-1})$ the point at which the current particle is before sampling $Y_i$. Similarly, we denote $w_{i-1}=w(S_{i-1})$ the statistical weight of this particle before sampling $Y_i$. 
From this point and until \cref{eq:conditional_on_P} we will work at a fixed entering state $S_{i-1}$ to try to express the $i^\text{th}$ term of \cref{eq:total_var_chrono}.
The term $\esp{R | S_i }$ represents the expected outcome of the result $R$ when the state is $S_i$. To express this term, we first need to write out explicitly how $R$ is computed depending on the $Y_i$:
\begin{equation}
    R = \frac{1}{N} \sum_{i=0}^\infty w(S_i) h\big(P(S_i)\big).
    \label{eq:expression_of_result}
\end{equation}

Let us decompose $\esp{R | S_i }$ into terms that depend on the current sampling event $Y_i$, and terms that do not depend on $Y_i$. From \cref{eq:expression_of_result}, the sum that yields the result $R$ can be artificially decomposed into the component of $R$ collected by the current particle and all its descendants, denoted $R_\text{curr}$ (also referred to as the contribution of the current particle), and the component of $R$ collected by the remaining portion of the simulated particles, denoted $R_\text{other}$. The quantity $R_\text{curr}$ can be further decomposed into two terms: the contribution collected before the current sampling $Y_i$, denoted $R_{\text{curr} <i}$; and the contribution that will be collected after the sampling $Y_i$, denoted $R_{\text{curr} \ge i}$. Summing all the contributions, we then have:
\begin{equation}
    \esp{R | S_i } = \esp{R_\text{other}|S_i} + \esp{R_{\text{curr} <i}|S_i} + \esp{R_{\text{curr} \ge i}|S_i}.
\label{eq:expectation_decomposed_current_other}
\end{equation}

Since all particles are independent, the sampling $Y_i$ that concerns the current particle cannot influence the expected contribution of all the other particles. This implies
\begin{equation}
    \esp{R_\text{other}|S_i} = \esp{R_\text{other}|S_{i-1}}.
\end{equation}
Similarly, the sampling of $Y_i$ does not influence the score already collected by the current particle, so that:
\begin{equation}
    \esp{R_{\text{curr} <i}|S_i} = \esp{R_{\text{curr} <i}|S_{i-1}}.
\end{equation}
The contribution of the current particle after the sampling $Y_i$ is provided by the terms that actually depend on $Y_i$. The contribution $R_{\text{curr}\ge i}$ collected after sampling $Y_i$, is the sum of all the contributions to $R$ of all the $K$ offspring of the current particle. For a single offspring (indexed by $k\in[1,K]$) that will be produced after $Y_i$ at point $P'_k$ with a weight $w'_k$, its expected score will be $w'_k M_1(P'_k)$, so its expected contribution to $R$ will be $w'_k M_1(P'_k)/N$. We can then compute the total expected score from all the offsprings as:
\begin{equation}
    \esp{R_{\text{curr} \ge i}|S_i}= \frac{1}{N}\sum_{k=1}^K w_k' M_1(P_k').
\label{eq:expectation_current_via_importance}
\end{equation}
We can then express the outgoing weights as $w'_k=m_kw$. Similarly, we express the outgoing importance as:
\begin{equation}
    M_1(P'_k) = M_1(P) \times \frac{M_1(P'_k)}{M_1(P)},
    \label{eq:express_rel_importance}
\end{equation}
and \cref{eq:expectation_decomposed_current_other} can be re-written as:
\begin{equation}
    \esp{R|S_i}= \frac{w M_1(P)}{N} \sum_{k=1}^K m_k \frac{M_1(P'_k)}{M_1(P)} + \esp{R_{\text{curr} <i}|S_{i-1}}+ \esp{R_\text{other}|S_{i-1}}.
    \label{eq:expectation_of_R}
\end{equation}

Now, if we take the variance of \cref{eq:expectation_of_R} conditioned to the initial state $S_{i-1}$, the terms $w$, $M_1(P)$, $\esp{R_{\text{curr} <i}|S_{i-1}}$ and $\esp{R_\text{other}|S_{i-1}}$ only depend on $S_{i-1}$. Thus, we have
\begin{equation}
    \var{\esp{R|S_i}|S_{i-1}} = \frac{w^2 M_1^2(P)}{N^2} \times \var{ \frac{\sum_{k=1}^K m_k M_1(P'_k)}{M_1(P)}}.
\label{eq:extrinsic_intrinsic_var}
\end{equation}
Note that in \cref{eq:express_rel_importance,eq:expectation_of_R,eq:extrinsic_intrinsic_var} we implicitly supposed that $M_1(P)$ was non-zero. This is where the hypothesis of a `positive' Monte-Carlo game is useful, since it implies\footnote{In fact, throughout this work, the `positivity' of the Monte-Carlo game is only used to ensure $M_1\ge 0$. Thus, the real condition for the variance decomposition to be valid is simply that $M_1\ge 0$.} that $M_1\ge 0$. Therefore, $M_1(P)$ can only be zero if all the outcomes of the process are null (as no canceling out is possible). In this case, the variance on the left-hand side of \cref{eq:extrinsic_intrinsic_var} is null.

To simplify \cref{eq:extrinsic_intrinsic_var}, we introduce the importance of the outcome of the sampling process $X_P$ as:
\begin{equation*}
    M_1(X_P) = h(P)+\sum_{k=1}^K m_k M_1(P'_k),
\end{equation*}
which corresponds to the definition in \cref{eq:def_importance_xp}. The importance of sampling $X_P$ satisfies\footnote{\Cref{eq:conservation_of_importance} expresses the `conservation of importance': it corresponds to the equation of the first moment. It can be obtained by several means, and a proof is given in \cref{sec:proof_law_of_total_expectation}.}
\begin{equation}
    \esp{M_1(X_P)}= M_1(P),
    \label{eq:conservation_of_importance}
\end{equation}
which allows us to rewrite \cref{eq:extrinsic_intrinsic_var} as:
\begin{equation}
    \var{\esp{R|S_i}|S_{i-1}} = \frac{w^2 M_1^2(P)}{N^2} \times \relvar{X_{P}},
\label{eq:extrinsic_intrinsic_var_2}
\end{equation}
where we used the shorthand notation $\relvar{X_P}$ introduced in \cref{eq:intrinsic_variance}. Again, if $M_1(P)=0$, the intrinsic variance $\relvar{X_P}$ is undefined, but the contribution of $X_P$ to the final relative variance will be zero. 

We have worked at fixed entering state $S_{i-1}$, and obtained \cref{eq:extrinsic_intrinsic_var_2}. We will now express the whole sum in \cref{eq:total_var_chrono} and re-arrange it by regrouping all the terms that have their current particle in the same initial state $P$. To do that we use the law of total expectation with the random variable $P_{i-1}$. For clarity, we will temporarily use the shorthand notation $Z_i := \var{\esp{R|S_i}|S_{i-1}}$. We have
\begin{align}
    \var{R} &= \sum_i \esp{\var{\esp{R|S_i}|S_{i-1}}}\nonumber\\
    &= \sum_i \esp{Z_i}\nonumber\\
    &= \sum_i \esp{\esp{Z_i|P_{i-1}}} \nonumber\\
    &= \sum_i \int_\Pi \esp{Z_i|P_{i-1}=P} f_{P_{i-1}}(P)dP,
    \label{eq:conditional_on_P}
\end{align}
where we used the law of total expectation in the next-to-last line, and we expressed the expected value as an integral on the values taken by $P_{i-1}$ in the last line, where $f_{P_{i-1}}(P)dP$ represents the probability that the current particle is in $dP$ around $P$ before the $i^\text{th}$ sampling.  Note that at discrete points such as $P_\text{source}$, where particles have a non-zero probability of being exactly at that point, the density $f_{P_{i-1}}$ uses the discrete measure and $f_{P_{i-1}}(P)$ represents the probability of being at $P$ before $Y_i$. We can then invert the sum over $i$ and the integral over phase space\footnote{As all the terms in the sums are positive, the inversion poses no problem, as long as the variance $\var{R}$ is well defined.}, replace the shorthand notation $Z_i$ that we introduced, and extract the terms that only depend on $P$ from the sum over $i$ :
\begin{align}
    \var{R}  &=  \int_\Pi \sum_i \esp{Z_i|P_{i-1}=P} f_{P_{i-1}}(P)dP\nonumber\\
    &= \frac{1}{N^2}\int_\Pi \sum_i f_{P_{i-1}}(P) \esp{ w_{i-1}^2 M_1^2(P_{i-1})\relvar{X_{P_{i-1}}}|P_{i-1}=P}dP\nonumber\\
    &= \frac{1}{N^2}\int_\Pi \left[ \sum_i f_{P_{i-1}}(P) \espw{w_{i-1}^2|P_{i-1}=P} \right] M_1^2(P)\relvar{X_P}dP.
\label{eq:invert_history_phase_space}
\end{align}

The sum over $i$ in \cref{eq:invert_history_phase_space} counts $w_{i-1}^2$ each time the current particle is in $dP$ around $P$. Note that $w_{i-1}$ is a random variable that depends on $S_{i-1}$, even if the point $P$ is fixed. We can then introduce rigorously the average number $n(P)dP$ of particles in $dP$ around $P$  by counting the occurrences of current particles being in $dP$ around $P$:
\begin{align}
    &n(P)dP = \sum_i f_{P_{i-1}}(P)dP\nonumber\\
    \Leftrightarrow\qquad& n(P)= \sum_i f_{P_{i-1}}(P).
    \label{eq:def_n}
\end{align}
Similarly, we introduce rigorously the matter density $\rho(P)$ as:
\begin{equation}
    \rho(P):=\frac{1}{N}\sum_i f_{P_{i-1}}(P)\espw{w_{i-1}|P_{i-1}=P} .
    \label{eq:def_matter_density}
\end{equation}
The average weight of particles at $P$ is defined by
\begin{equation}
    \espw{w(P)}:= \frac{\rho(P)}{n(P)/N},
    \label{eq:def_esp_w}
\end{equation}
which corresponds to the definition in \cref{eq:def_avg_weight_from_rho}. Note that, even though $w(P)$ is not exactly a random variable (as in some realisations of the game, no particles might land on point $P$ for example), we still use the notation of $\mathbb{E}$ for expected value as $\espw{w(P)}$ can be reformulated as an expected value in another probability space. Our original probability space has a sample space composed of the different possible realizations of our Monte Carlo game.
If our game is feasible, then almost surely all particles are dead in a finite time and $\sum_i f_{P_{i-1}}$ form a finite measure (called Campbell's measure) over $\Pi/\{P_\text{death}\}$. Therefore, by renormalizing that measure, we can see that there is a new probability space, often called Palm's space, where the sample space is a particle in a possible game at an instant $i$, at a point $P$, with a weight $w$. Instead of sampling a game, we sample a possible particle state within all the particles in all the possible games. Using that new probability space one can express the random variable $w$ that yields the weight of the particle sampled at the time it is sampled. $\espw{w(P)}$ defined according to \cref{eq:def_esp_w} corresponds to the expected value of $w$ conditioned on the fact that the particle lands on point $P$. The curious reader can learn more about Palm's theory that lies behind this reasoning by reading Chapter 3 of Ref~\citep{baccelli_random_2024}. 

Similarly, the average square weight of particles at point $P$ is defined by counting $w_{i-1}^2$ each time the particle is at point $P$, and dividing by the average number $n(P)$ of particles:
\begin{equation}
    \espw{w^2(P)} := \frac{\sum_i f_{P_{i-1}}(P)\espw{w_{i-1}^2|P_{i-1}=P}}{n(P)}.
\end{equation}
It also corresponds to the expected value of $w$ conditioned on landing on $P$ in our modified probability space. For a point $P$ that sees  no particles ($n(P)=0$), which means that it is `unreachable' from the source, the quantities $\espw{w(P)}$ and $\espw{w^2(P)}$ are undefined. In this case, \cref{eq:def_n} implies that $f_{P_{i-1}}(P)=0\;, \forall i$, and \cref{eq:invert_history_phase_space} implies that the integrand is null at that point $P$. 

From \cref{eq:invert_history_phase_space} we have then
\begin{align}
    \var{R} &= \frac{1}{N^2} \int_\Pi \espw{w^2(P)} n(P) M_1^2(P)\relvar{X_P}dP.
    \label{eq:preliminary_var_decomp}
\end{align}
Now, we want to replace the terms in the formula with others that are more practical to obtain or more directly affected by variance-reduction techniques. We decompose
\begin{equation}
    \espw{w^2(P)} = \espw{w(P)}^2+ \varw{w(P)} = \espw{w(P)}^2 \left(1+\relvarw{w(P)} \right),
\end{equation}
where $\relvarw{w(P)}=\varw{w(P)}/\espw{w(P)}^2$. By using the sampling intensity $I_s$ defined in \cref{eq:sampling_intensity} and the relative contribution density $c_r(P)$ defined in \cref{eq:def_rel_contrib}, we rearrange the terms in \cref{eq:preliminary_var_decomp} to obtain the final variance-decomposition formula:
\begin{equation*}
    \relvar{R} = \frac{1}{N}\int_\Pi c_r(P)\frac{1+\relvarw{w(P)}}{I_s(P)}\relvar{X_P}dP,
\end{equation*}
which corresponds to \cref{eq:decomp_var}.

For the points $P$ where the importance $M_1(P)=0$ is null, the relative contribution $c_r(P)$ and the whole integrand is null, although $I_s(P)$ and $\relvar{X_P}$ are undefined. For the points $P$ where the number of particles is zero (unreachable points), the relative contribution $c_r(P)$ is null, and the integrand is null, although $\relvarw{w(P)}$ and $I_s(P)$ are undefined. 

\subsection{The law of total expectation}
\label{sec:proof_law_of_total_expectation}

Since the variance-decomposition formula was obtained using the law of total variance on the result $R$, one might be interested in similarly applying the law of total expectation. The law of total expectation states that, given two random variables $A$ and $B$, if $A$ has an expected value, then (see e.g.~Section 4.4 of Ref.~\citep{casella_statistical_2024})
\begin{equation}
    \label{eq:law_of_total_expectation}
    \esp{A}=\esp{\esp{A|B}}.
\end{equation}
Using again our setup for the response $R$ and the different samplings $Y_i$, one can write an ensemble of equations:
\begin{equation}
    \forall i \in \mathbb{N}^*,\; \esp{R|S_{i-1}}=\esp{\esp{R|S_{i-1},Y_i}}.
    \label{eq:law_of_total_expectation_on_R}
\end{equation}
The presence of a system of equations is in stark contrast with the law of total variance, which yielded a single equation \cref{eq:total_var_chrono}.

Again, given a fixed state $S_{i-1}$, we define the `current particle' being at point $P$, and we then decompose $R$ into $R_\text{other}$, $R_{\text{curr}<i}$ and $R_{\text{curr}\ge i}$, similarly to what done in \cref{eq:expectation_decomposed_current_other}. We therefore obtain:

\begin{align}
    &\esp{R_\text{other}|S_{i-1}}+\esp{R_{\text{curr}<i}|S_{i-1}}+\esp{R_{\text{curr}\ge i}|S_{i-1}} \nonumber\\
    &= \esp{\esp{R_\text{other}|S_i}+\esp{\esp{R_{\text{curr}<i}|S_i}}+\esp{\esp{R_{\text{curr}\ge i}|S_i}}}.
    \label{eq:law_of_total_exp_decomposed_R}
\end{align}
Since $R_\text{other}$ does not depend on the sampling $Y_i$, we have $\esp{R_\text{other}|S_{i-1}}=\esp{R_\text{other}|S_i}$, and both terms cancel out on each side of \cref{eq:law_of_total_exp_decomposed_R}. The same argument applies to $R_{\text{curr}<i}$.

The term $\esp{R_{\text{curr}\ge i}|S_{i-1}}$ describes the total future contribution to $R$ collected by the current particle (currently at point $P$) and its offsprings. This corresponds to $w M_1(P)$ if we remove the score collected at point $P$, which occurred before the sampling $Y_i$)=:
\begin{equation}
    \esp{R_{\text{curr}\ge i}|S_{i-1}}= \frac{w}{N}\left(M_1(P) - h(P)\right).
\end{equation}
Using \cref{eq:expectation_current_via_importance} to replace $\esp{R_{\text{curr}\ge i}|S_i}$ in \cref{eq:law_of_total_exp_decomposed_R}, we obtain
\begin{equation}
    \frac{w}{N}\left(M_1(P) - h(P)\right) = \frac{1}{N}\sum_{k=1}^K w_k' M_1(P_k').
\end{equation}
Finally, by introducing the weight multipliers $m_k$ to factorize the weight $w$, we get:
\begin{equation}
    \esp{M_1(P)}=\esp{h(P)+\frac{1}{N}\sum_{k=1}^K w_k' M_1(P_k')}=\esp{M_1(X_P)},
\end{equation}
which proves \cref{eq:conservation_of_importance}.

\setcounter{equation}{0}
\section{Analysis of population-control procedures}
\label{sec:proof_split_roul}

In this appendix, we prove \cref{eq:var_pop_control} that was introduced in \cref{sec:pop_control}. We use the notation of \cref{sec:pop_control}.

The importance of the sampling $X_{P_w}$ that applies population control on weight $w$ at point $P$ is (from \cref{eq:def_importance_xp} considering that $h(P_w)=0$ as we do not score on the entering point of the weight control procedure) :
\begin{equation}
    M_1(X_{P_w}) = \sum_{k=1}^K m_k M_1(P'_{k}),
\end{equation}
where the total number of offsprings $K$ is either $\lfloor f \rfloor$ or $\lfloor f+1 \rfloor$. The weight multiplier $m_k$ is always $1/f$, and $P'_{k}$ is simply $P_\text{out}$, i.e.~the point in phase space corresponding to $P$ but right after the population control. Therefore, $M_1(X_{P_w})$ is a binary variable that takes the value $\lfloor f \rfloor M_1(P_\text{out})/f$ with probability $\lfloor f+1\rfloor -f$, and $\lfloor f +1 \rfloor M_1(P_\text{out})/f$ with probability $f-\lfloor f \rfloor$. This process is unbiased, since we have:
\begin{align}
    M_1(P_w) &=\esp{M_1(X_{P_w})} \nonumber \\
    &=\lfloor f \rfloor \frac{M_1(P_\text{out})}{f} \left(\lfloor f+1\rfloor -f \right) + \lfloor f +1 \rfloor\frac{M_1(P_\text{out})}{f} \left(f-\lfloor f \rfloor\right) \nonumber \\
    &= M_1(P_\text{out}),
\end{align}
i.e. the weight control process has no influence on the expected score of a particle. The relative variance of $M_1(X_{P_w})$ is:
\begin{align}
    \relvar{M_1(X_{P_w})} &= \var{\frac{M_1(X_{P_w})}{\esp{M_1(X_{P_w})}}} \nonumber \\
    &=\left( \frac{\lfloor f+1 \rfloor}{f}- \frac{\lfloor f \rfloor }{f}\right)^2(f-\lfloor f \rfloor)(\lfloor f+1\rfloor-f) \nonumber \\
    &= \frac{(f-\lfloor f \rfloor)(\lfloor f+1\rfloor-f)}{f^2}.
\end{align}

Now, let us derive the relative contribution of the fraction of particles at point $P$ with weight $dw$ around $w$. The total relative contribution of all particles (with all weights) in $dP$ around $P$ is $c_r(P)dP$. Using \cref{eq:def_avg_weight_from_rho,eq:def_rel_contrib}, the fraction of this contribution at weight $dw$ around $w$ is then:
\begin{equation}
    c_r(P,w)dPdw = c_r(P)dP \times \frac{w n(P,w)dw}{n(P)\espw{w(P)}}.
\end{equation}

The sampling intensity can be calculated as the unitary number of particles divided by the relative contribution, namely
\begin{align}
    I_s(P,w) &= \frac{n(P,w)}{Nc_r(P,w)}\nonumber\\
    &= \frac{n(P)\espw{w(P)}}{N c_r(P) w}\nonumber\\
    &= I_s(P)\frac{\espw{w(P)}}{w}.
\end{align}

Finally, the relative variance of the weights is zero, since we are only considering the particles with weight $w$. We will now integrate the variance-decomposition terms from all the different weights at point $P$, which yields:
\begin{align*}
    \relvar{\text{Pop. cont. in }dP\text{ around } P} &= \int_{w=0}^\infty c_r(P,w) \frac{1+ \relvarw{w(P_w)}}{I_s(P,w)}\relvar{X_{P_w}}dPdw
    \\&=\frac{c_r(P)dP}{I_s(P)} \int_{w=0}^\infty \frac{n(P,w)}{n(P)}\frac{(w_t(w))^2}{\espw{w(P)}^2} (f-\lfloor f \rfloor)(\lfloor f+1\rfloor-f) dw.
\end{align*}
This is precisely the result stated in \cref{eq:var_pop_control}.The physical interpretation of this equation is provided in \cref{sec:pop_control}.

\newpage
\bibliographystyle{tfcse}
\bibliography{biblio}

\end{document}